\documentclass[
preprint,
superscriptaddress,
amsmath,amssymb,
aps,
prl,
10pt,
twocolumn,
]{revtex4-2}
\usepackage{float}
\usepackage{graphicx}
\usepackage{dcolumn}
\usepackage{bm}
\usepackage[dvipsnames]{xcolor}
\usepackage[colorlinks,citecolor=blue]{hyperref}
\usepackage{setspace}
\usepackage{indentfirst}
\usepackage{braket}
\providecommand{\keywords}[1]
{
\small\textbf{\textit{Keywords:})}}
\begin{document}
	
	\title{Non-orthogonal cavity modes near exceptional points in the far field}
	
	\author{Jingnan Yang}
	\affiliation{State Key Laboratory for Mesoscopic Physics and Frontiers Science Center for Nano-optoelectronics, School of Physics, Peking University, 100871 Beijing, China}
	\author{Shushu Shi}
	\author{Sai Yan}
	\author{Rui Zhu}
	\affiliation{Beijing National Laboratory for Condensed Matter Physics, Institute of Physics, Chinese Academy of Sciences, Beijing 100190, China}
	\author{Xiaoming Zhao}
	\affiliation{Department of Physics and Institute of Theoretical Physics, University of Science and Technology Beijing, Beijing 100083, China}
	\author{Yi Qin}
	\affiliation{Guangdong Provincial Key Laboratory of Quantum Metrology and Sensing $\&$ School of Physics and Astronomy, Sun Yat-Sen University (Zhuhai Campus), Zhuhai 519082, China}
	\author{Bowen Fu}
	\affiliation{State Key Laboratory for Mesoscopic Physics and Frontiers Science Center for Nano-optoelectronics, School of Physics, Peking University, 100871 Beijing, China}
	\author{Xiqing Chen}
	\affiliation{State Key Laboratory for Mesoscopic Physics and Frontiers Science Center for Nano-optoelectronics, School of Physics, Peking University, 100871 Beijing, China}
	\author{Hancong Li}
	\affiliation{State Key Laboratory for Mesoscopic Physics and Frontiers Science Center for Nano-optoelectronics, School of Physics, Peking University, 100871 Beijing, China}
	\author{Zhanchun Zuo}
	\author{Kuijuan Jin}
	\affiliation{Beijing National Laboratory for Condensed Matter Physics, Institute of Physics, Chinese Academy of Sciences, Beijing 100190, China}
	\author{Qihuang Gong}
	\affiliation{State Key Laboratory for Mesoscopic Physics and Frontiers Science Center for Nano-optoelectronics, School of Physics, Peking University, 100871 Beijing, China}
	\author{Xiulai Xu}
	\email{xlxu@pku.edu.cn}
	\affiliation{State Key Laboratory for Mesoscopic Physics and Frontiers Science Center for Nano-optoelectronics, School of Physics, Peking University, 100871 Beijing, China}
	\affiliation{Peking University Yangtze Delta Institute of Optoelectronics, Nantong, Jiangsu 226010, China}

   \begin{abstract}
		\textbf{ Non-orthogonal eigenstates are a fundamental feature of non-Hermitian systems and are accompanied by the emergence of nontrivial features. However, the platforms to explore non-Hermitian mode couplings mainly measure near-field effects, and the far-field behaviour remain mostly unexplored. Here, we study how a microcavity with non-Hermitian mode coupling exhibits eigenstate non-orthogonality by investigating the spatial field and the far-field polarization of cavity modes. The non-Hermiticity  arises from asymmetric backscattering, which is controlled by integrating two scatterers of different size and location into a microdisk. We observe that the spatial field overlaps of two modes increases abruptly to its maximum value, whilst different far-field elliptical polarizations of two modes coalesce when approaching an exceptional point. We demonstrate such features experimentally by measuring the far-field polarization from the fabricated microdisks. Our work reveals the non-orthogonality in the far-field degree of freedom, and the integrability of the microdisks paves a way to integrate more non-Hermitian optical properties into nanophotonic systems.
}

	\end{abstract}
	\maketitle
	
	\section{Introduction}
	Hermiticity in quantum mechanics guarantees orthogonality and completeness of eigenstates in a Hermitian system.
	In contrast, eigenstates of a non-Hermitian system are usually non-orthogonal.
	The eigenstate non-orthogonality is the fundamental and distinct feature to reveal the non-Hermiticity \cite{doi:10.1080/00018732.2021.1876991, Rotter_2009}, and provides the basis for key non-Hermitian phenomena such as the exceptional point (EP). In other words, the eigenstates and eigenvalues simultaneously coalesce at peculiar spectral singularities \cite{Gao2015, PhysRevE.69.056216, PhysRevLett.99.173003, PhysRevLett.121.085702}.
	EP enables many counter-intuitive phenomena including the chiral absorption \cite{Soleymani2022,Wang2020} and the strong modified spontaneous emission \cite{Lu:22, Kim2021}.
	Therefore, degrees of freedom that exhibit the eigenstate non-orthogonality are highly desired for functions of non-Hermitian quantum devices \cite{doi:10.1073/pnas.1603318113}, such as discriminating non-Hermitian quantum states \cite{PhysRevA.106.022438} and exploring non-Hermitian topological physics \cite{doi:10.1126/science.aaz8727,PhysRevX.6.021007}.
	
	Recently, optical systems have been attracting increasing interests in studying non-Hermitian physics \cite{doi:10.1126/science.aar7709, Feng2017, PhysRevLett.118.093002, PhysRevLett.121.093901, PhysRevLett.128.223903}, such as the whispering gallery mode (WGM) microcavities with asymmetric backscattering \cite{PhysRevA.84.063828, doi:10.1073/pnas.1603318113}, parity-time ($PT$) or anti-$PT$ symmetry \cite{Feng2017, PhysRevLett.124.053901}.
	Such microcavities provide a platform to investigate the non-orthogonal modes in different degrees of freedom, such as propagation direction \cite{doi:10.1073/pnas.1603318113, Biasi:22}, resonance frequency \cite{Chen2017}, intensity and beam dynamics \cite{PhysRevLett.100.103904, 1070064}.
	Another degree of freedom, far-field polarization, is correlated to the symmetry and the orthogonality of near field \cite{Vukovi2002OptimizationOT, Chalcraft:11, Xiong2018} and plays key roles in nanophotonics, such as single-photon sources \cite{Wang2019}, lasers \cite{He2016}, nonlinear frequency conversion \cite{10.1002/lpor.202200183} and topological optics \cite{10.3389/fphy.2022.862962} and structured light \cite{Forbes2021}.
	However, the measurement of  WGM microcavities with non-Hermitian mode coupling or $PT$ symmetry mainly relies on the near-field fiber coupling \cite{doi:10.1073/pnas.1603318113, Chen2017}.
	As such, far-field polarization features of such WGM microcavities remain unexplored.

	We use an ensemble of quantum dots as internal broad-band light sources to enable the far-field excitation \cite{doi:10.1126/science.1258479},
	and use integrated internal scatterers to reduce the effects of disturbances from using external scatterers in the far-field collection.
	By integrating internal scatterers such as air cuts, bulges or holes, non-Hermitian mode coupling can be obtained in the WGM microcavities \cite{Kim:14,PhysRevA.93.033809,Yang:21, PhysRevA.108.L041501}.
	Furthermore, compared to traditional cavities with non-Hermitian mode coupling realized by external scatterers or gain/loss, the integrated internal scatterers enable small-size designs with small mode volumes and large-scale on-chip integration.
	These two features provide the basis for studying non-Hermitian cavity quantum electrodynamics (QED) and quantum optical devices \cite{PhysRevResearch.3.043096, Lu:22, PhysRevResearch.2.023375, Pick:17, PhysRevLett.117.107402}.
	
	Here we study the polarization of far-field photoluminescence (PL) spectra from cavity modes in microcavities with non-Hermitian mode coupling around an EP.
	The non-Hermiticity in mode coupling is realized by the asymmetric backscattering from two internal weak scatterers of different size and location, and the mode non-orthogonality is further controlled by the relative angle between the two scatterers.
	We reveal the angle dependence of the spatial field distribution and the far field polarization as non-Hermitian features of the system.
	Approaching an EP, we observe that the spatial field overlap of two modes increases abruptly, and the far-field polarizations of two modes are nearly identical, corresponding to coalescent eigenstates at an EP.
	These polarization features are further demonstrated by experimental measurements of fabricated microdisks.
	Since our microdisks with internal weak scatterers can be integrated on-chip, the non-orthogonality based near- and far-field features enable further applications to integrate the non-Hermitian optical properties into nanophotonic systems.
	
	\section{Results}
	\subsection{Theoretical analysis}
		\begin{figure*}
		\centering
		\includegraphics[scale=0.22]{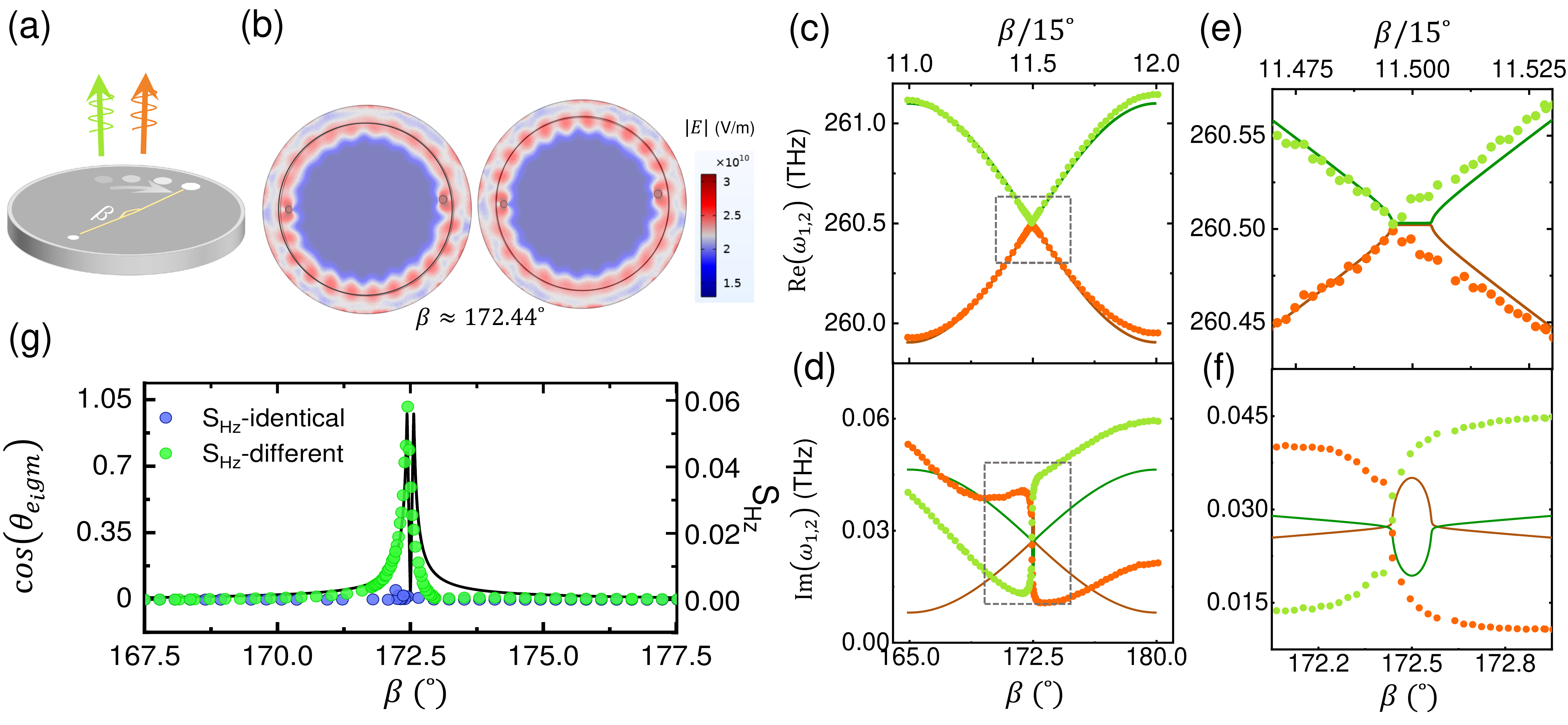}
		\caption{Simulation and theoretical results for a microcavity with two scatterers. (a) Illustration of a microdisk integrated with two holes. (b) Simulated near-field $\lvert E \rvert$ distribution of two modes close to an EP with $\beta\approx172.44^\circ$. Here $r_{1}=0.037285$ $\mu$m, $r_{2}=0.03203$ $\mu$m, $d_{1}=0.73312$ $\mu$m, $d_{2}=0.74995$ $\mu$m. (c)-(d) Theoretical (solid lines) and simulated (dots) complex frequencies of two modes respectively colored in green and orange with their real parts (c) and imaginary parts (d). (e)-(f) Magnified regions labeled by dashed squares in (c) and (d). (g) $\cos(\theta_{eigm})$ (solid black line) and $S_{H_{z}}$ varying with $\beta$ for two different holes (green dots) and two identical holes (blue dots).}
		\label{p1}
	\end{figure*}

	Left and right eigenstates, $\ket{\psi_{i}^{L}}$ and $\ket{\psi_{i}^{R}}$, correspond to Hamiltonians $H$ and $H^{\dagger}$, respectively \cite{PhysRevB.102.245147}.
	In contrast to a Hermitian system, in a non-Hermitian system we have $H \neq H^{\dagger}$.
	Therefore, the non-Hermitian system follows the bi-orthogonality $\langle\psi_{i}^{L}\ket{\psi_{j}^{R}}= \delta_{i,j}$, which is very challenging to be directly experimentally obtained.
	However, the self-orthogonality $\langle\psi_{i}^{R}\ket{\psi_{j}^{R}}$ and $\langle\psi_{i}^{L}\ket{\psi_{j}^{L}}$, which follows $\delta_{i,j}$ for a Hermitian system, do not vanish for $i\neq j$ for a non-Hermitian system.
	This is the intrinsic difference between Hermitian and non-Hermitian systems.
	The non-orthogonality $\langle\psi_{i}^{R}\ket{\psi_{j}^{R}}$ supports non-trivial phenomena such as chirality \cite{doi:10.1073/pnas.1603318113} and topology \cite{RevModPhys.93.015005}, therefore, it is one focus of theoretical and experimental investigations on non-Hermitian systems.
	
	Non-Hermitian optical systems with non-Hermitian mode coupling have been achieved by the asymmetric backscattering between clockwise (CW) and counter-clockwise (CCW) traveling-wave modes in a WGM microcavity.
	These modes are quasinormal modes since the optical microcavity is non-conservative with energy loss \cite{10.1002/lpor.201700113, doi:10.1021/acsphotonics.0c00014, Lalanne:19}.
	We only focus on well confined transverse electric (TE) modes in the first radial order with a high quality (Q) factors \cite{Srinivasan:06} and dominating $E_{x}$, $E_{y}$ and $H_{z}$ components, which can be easily experimentally observed in far-field photoluminescence (PL) spectra from microcavities with embedded self-assembled quantum dots as excitation sources\cite{Yang:21, Silva:08, 4749336, PhysRevB.63.233306,PhysRevLett.122.087401}.
	Here the TE modes are labeled by $TE_{1,m}$ with the azimuthal mode number $m$. The cavity mode function follows $\psi_{CW} \propto e^{-im\phi}$ and $\psi_{CCW} \propto e^{im\phi}$ thus has $2m$ antinodes along the cavity perimeter.
	$\phi$ is the azimuthal angle in the polar coordinate.
	
	WGMs are very sensitive to nano-scale perturbations \cite{doi:10.1073/pnas.1408453111, Yang2020, Kippenberg:02, Li:12}.
	These perturbations act as Rayleigh scatterers that introduce the backscattering coupling between CW and CCW modes, which are degenerate without scatterers.
	The non-Hermiticity in WGM mode coupling can be realized and modulated by the asymmetric backscattering from weak scatterers \cite{PhysRevA.84.063828} or strong scatterers \cite{PhysRevA.108.L041501}.
	Here we consider a WGM cavity with two weak scatterers and use a two-mode approximation based Hamiltonian picture to analyze how the non-Hermiticity is controlled by the scatterers.
	The corresponding Hamiltonian using the traveling-wave basis $\ket{\psi_{CW}^{R}}$ and $\ket{\psi_{CCW}^{R}}$ is written as
	\begin{equation}
		H = \left(\begin{array}{cc}
			\omega_{0} + \omega^{'} & A + Be^{-i2m\beta} \\
			A + Be^{i2m\beta} & \omega_{0} + \omega^{'}
			\label{matrix1}
		\end{array}\right).
	\end{equation}
	The terms $A$ and $B$ are complex because the scatterers induce a frequency shift and extra loss to the cavity modes, and they describe the backscattering effect from the two different weak scatterers, respectively. For more details about the backscattering, see Supplementary note 3 in supplementary materials (SMs).
	$\omega_{0}$ is the complex eigenvalues for the unperturbed microcavity, $\omega^{'}$ represents the complex frequency shift resulting from the two scatterers, and $\beta$ is the relative angle between the two scatterers, as illustrated in Fig. \ref{p1}(a).
	Eigenvalues and eigenstates can be obtained respectively as
	\begin{equation}
		\omega_{1,2} = 	\omega_{0} + \omega^{'} \pm \sqrt{A^{2} + B^{2}+2AB\cos(2m\beta)}\label{matrix2},
	\end{equation}
	\begin{equation}
		\ket{\psi_{1,2}^{R}}=M\ket{\psi_{CW}^{R}} \pm N \ket{\psi_{CCW}^{R}}.
	\end{equation}
	$M=\sqrt{A + Be^{-i2m\beta}}/\sqrt{2A+2B\rm{cos}(2m\beta)}$ and $N=\sqrt{A + Be^{i2m\beta}}/\sqrt{2A+2B\cos(2m\beta)}$. The eigenfrequencies are periodically modulated by $\beta$ with period $P=180^{\circ} / m=15^{\circ}$ for $m=12$. To quantify the eigenstate non-orthogonality, we introduce $ \cos(\theta_{eigm}) =\langle{\psi_{1}^{R}}\ket{\psi_{2}^{R}}$,  where we call $\ket{\psi_{1}^{R}}$ and $\ket{\psi_{2}^{R}}$ orthogonal (coalescent) if $ \cos(\theta_{eigm})=0$ ($\cos(\theta_{eigm})=1$).
	
	When the two weak scatterers differ ($A \neq B$), eigenstates exhibit non-orthogonality.
	In particular, EPs can be realized when one of the non-diagonal terms of $H$ in Eq.~\eqref{matrix1} vanishes near $\beta= (L + 1/2)\times15^{\circ}$ ($L$ is an integer) and $H$ becomes defective \cite{PhysRevA.84.063828, PhysRevA.22.618}.
	To further explore the non-Hermiticity, we calculate the dependence of $\omega_{1,2}$ and $\rm{cos}(\theta_{eigm})$ on $\beta$ for $m=12$, and present the results in Fig.~\ref{p1}(c)-(g).
	The values for $A$, $B$ and $\omega_{0} + \omega^{'}$ are deduced from complex eigenfrequencies in simulation results, for details see \textit{Theoretical calculations of eigenvalues} in Methods. From eigenvalues in Eq.~\eqref{matrix2}, it can be inferred that non-Hermitian degeneracy happens when $A^2 +B^2+2AB\cos(2m\beta)=0 $, which is possible when $\vert A \vert \approx \vert B \vert$, $\cos(2m\beta)\approx-1$ and $\sin(2m\beta)\approx0$. For every $\beta= (L + 1/2)\times15^{\circ}$, there are two different $\beta$ that can satisfy this requirement, but correspond to different eigenmodes.
	As shown in Fig. \ref{p1}(c)-(d), resonance frequencies (c) and decay rates (d) of two modes become degenerate symmetrically around $\beta=172.5^{\circ}$ with two EPs.
		\begin{figure*}
		\centering
		\includegraphics[scale=0.32]{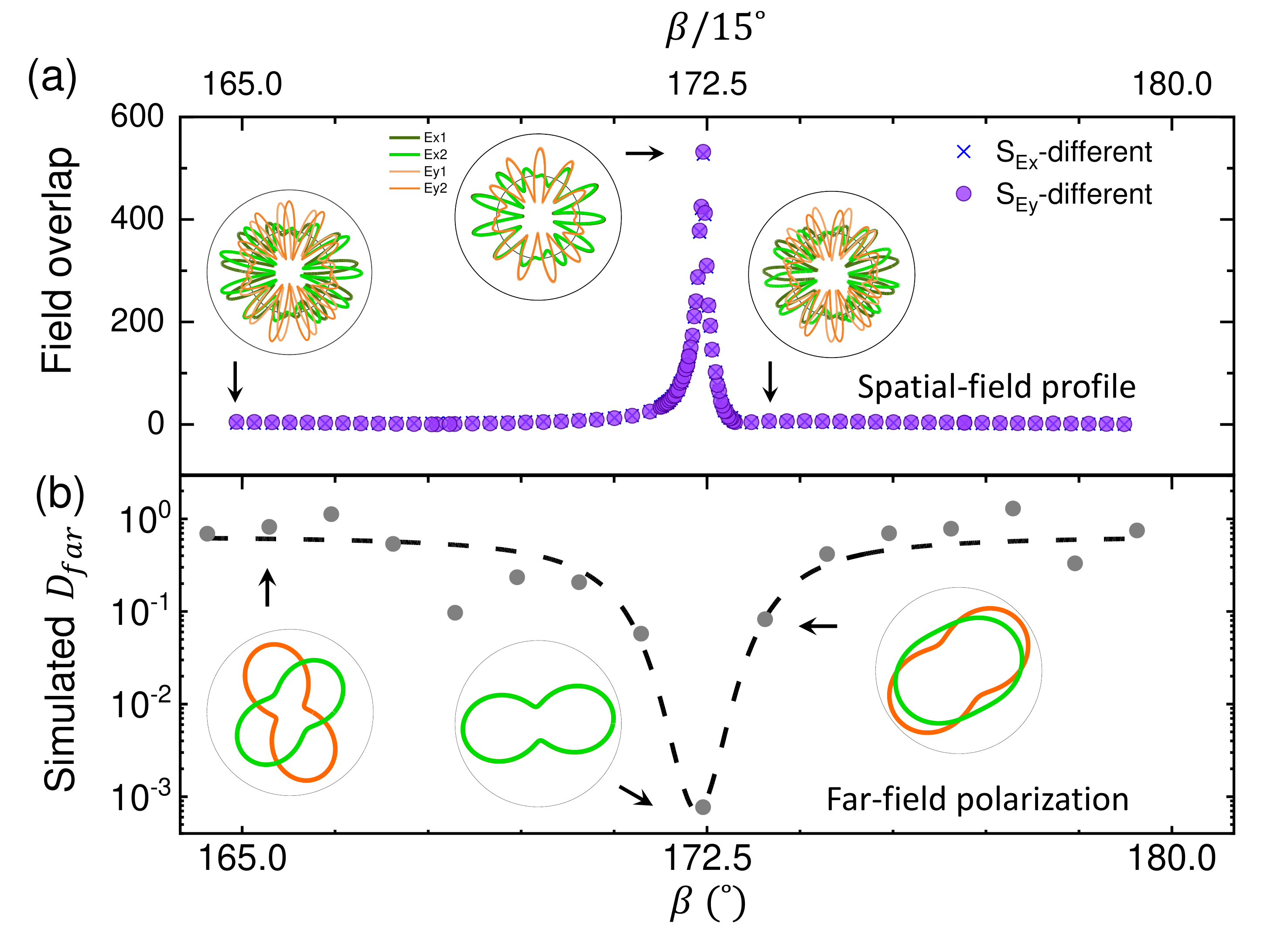}
		\caption{Simulation results for the spatial field overlap and the far-field polarization. (a) $S_{E_{x}}$ (blue crosses) and $S_{E_{y}}$ (purple dots) of two modes for two different scatterers where the maximum value corresponds to high coalescence while the minimum values correspond to high orthogonality. Insets: $E_{x}$ (green lines) and $E_{y}$ (orange lines) of two modes (subscript 1 and 2) with distributed along a black circle inside the microdisks for three different $\beta$ as shown in Fig. \ref{p1}(b).  (b) Far field polarization difference $D_{far}$ between two modes labeled by gray dots. The dashed gray line is the Lorentz fitting curve. Insets: polar maps of normalized polarization ellipses of two modes colored in orange and green for different $\beta$ by setting the ellipse area as 1.}
		\label{p2}
	\end{figure*}

	\subsection{Simulation results}
	\begin{figure*}
		\centering
		\includegraphics[scale=0.25]{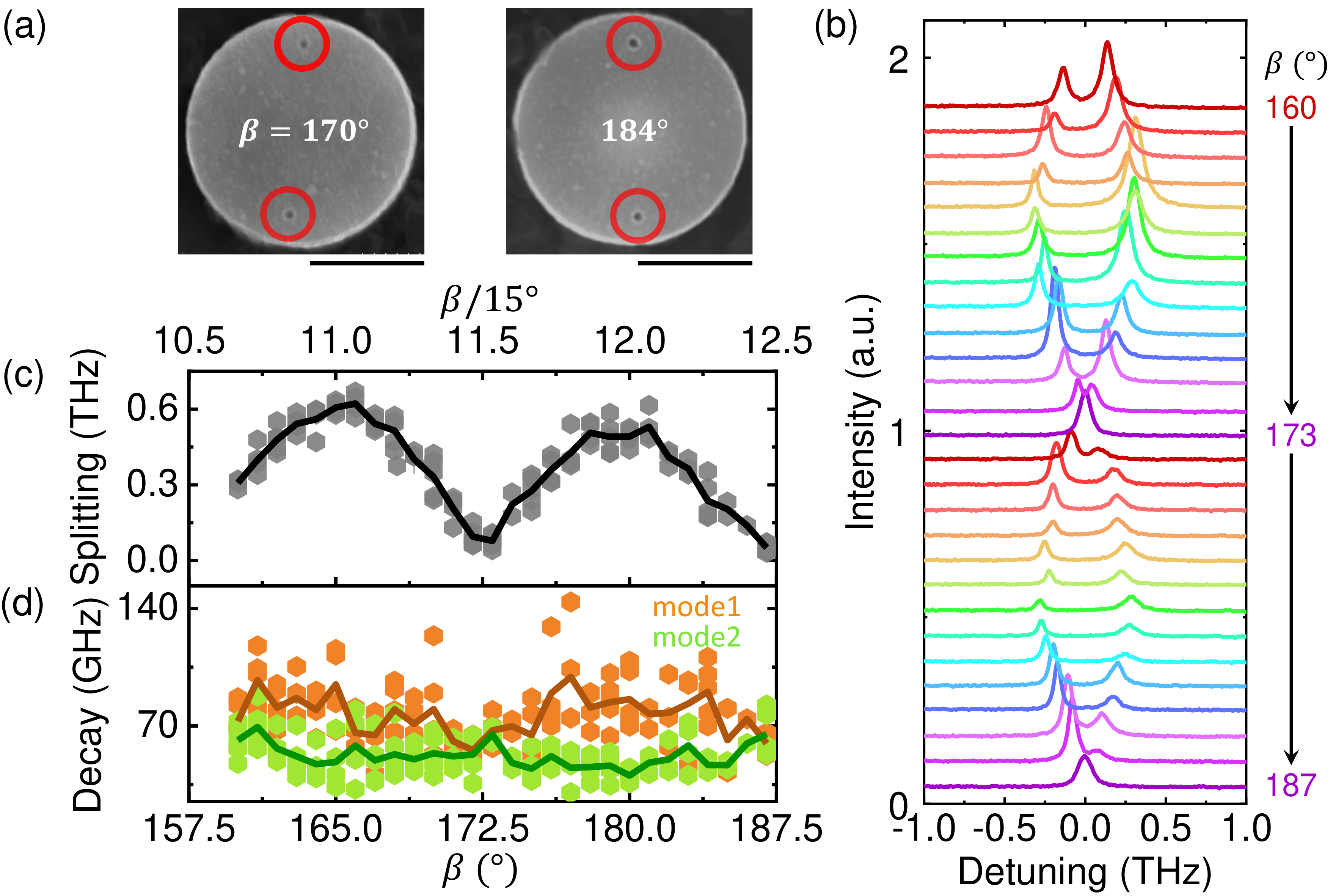}
		\caption{Experimental far-field PL spectra from fabricated microcavities with different $\beta$. (a) Scanning electron microscope (SEM) images of fabricated microdisks with two holes labeled by red circles for $\beta=170^{\circ}$ and $\beta=184^{\circ}$. The scale bars are 1 $\mu$m. (b) PL spectra of cavity modes for $\beta$ from $160^{\circ}$ to $187^{\circ}$. Here the center resonance frequency of each doublet is set as 0 THz for a better demonstration.  (c)-(d) Statistics of fitting results of resonance frequency splitting (c) and decay rates (d) of two modes.}
		\label{p3}
	\end{figure*}
	We implement numerical simulations using the finite element method.
	To control the non-Hermiticity, cylinder air holes are set in the microdisks with a thickness of  250~nm and a radius of 1 $\mu$m as shown in Fig. \ref{p1}(a).
	The two scatterers are defined by the radius $r_{1,2}$, the distance from the microdisk center $d_{1,2}$ and the relative angle $\beta$.
	We use $r_{1,2}=37.285,\ 32.03$ nm and $d_{1,2}=733.12,\ 749.95$ nm, which follows $r_{1}>r_{2}$ and $d_{1}<d_{2}$ to achieve $\vert A \vert \approx \vert B \vert$ for an EP \cite{PhysRevA.84.063828}.
	Details about the designs and the simulation methods are presented in Methods and supplementary note 1 in SMs.
	
	We focus on the spatial field distribution and far-field polarization of the two modes.
	At an EP, two modes propagate as traveling waves and coalesce into one mode with the same complex eigenfrequencies.
	As shown in Fig. \ref{p1}(b), the contrast between near-field antinodes and nodes for $\beta\approx172.44^{\circ}$ becomes vague, exhibiting features similar to traveling wave. More obvious features like traveling wave rather than standing wave can be seen in Fig. S1(j) in SMs.
	Meanwhile, two modes have almost the same complex eigenvalues as shown in Fig. \ref{p1}(e)-(f).
	$E_{x}$ and $E_{y}$ components of two modes distributed along a black circle on the microdisk surface as shown in Fig. \ref{p1}(b) highly coincide at $\beta\approx172.44^{\circ}$ as shown in Fig. \ref{p2}(a).
	In contrast, when away from the EP (e.g., $\beta \approx165^{\circ}$), an obvious spatial phase difference exists between $E_{x}$ and $E_{y}$ components of the two modes, as shown in Fig. \ref{p2}(a).
	To describe the spatial field overlap of two modes, we define $S_{k}$ ($k=E_{x}, E_{y}$ or $H_{z})$ to quantify the non-orthogonality.
	Take $H_{z}$ for example, we have
	\begin{equation}
		S_{H_{z}}=\Big\vert {\int{H_{z1}}^{*}{H_{z2}}dV}\Big\vert,
	\end{equation}
	where $H_{zi}$ is the normalized field extracted from the simulations.
	The normalization considers the field in the perfectly matched layer due to the non-Hermiticity of the quasinormal modes \cite{10.1002/lpor.201700113}.
	For details of the normalization, see \textit{Normalization of simulated spatial field} in Methods.
	Figure \ref{p1}(g) shows $\cos(\theta_{eigm})$ from the two-mode picture Eq. \ref{matrix1} and $S_{H_{z}}$ for two different holes (labeled by green circles) from the simulation with varying $\beta$.
	Both two parameters, which quantify the non-orthogonality, increase abruptly when approaching the EP and reach their the maximum values at the EP.
	Similar changes also apply to $S_{E_{x}}$ (blue crosses) and $S_{E_{y}}$ (purple dots) as shown in Fig. \ref{p2}(a), which exhibit the same trend but in a larger ratio versus $\beta$ compared to $S_{H_{z}}$.
	By comparison, such high non-orthogonality is not observed for two identical holes as shown by the results in Fig. \ref{p1}(g) (labeled by blue circles) and Fig. S2 (c) and (d) in Supplementary note 2 in SMs.
	The near-field distributions are strongly molded by the holes because of their direct effects on the eigenstate non-orthogonality. By comparison, the extra energy loss of the cavity caused by the scatterers is relatively weakly molded because it is affected by the spatial overlap between scatterers and mode field.
	
	Note that $\cos(\theta_{eigm})$ in Fig. \ref{p1}(g) indicates two EPs while $S_{H_{z}}$ only responds to one EP at $\beta\approx172.44^{\circ}$.
	Another EP is expected to appear at $\beta\approx172.56^{\circ}$ based on the two-mode picture (Eq. \ref{matrix1}).
	The deviation lies in the imperfection of the two-mode approximation model we use instead of the imperfection of the simulation mesh size, for example, only considering TE modes with a fixed $m$ is not a good assumption for internal scatterers \cite{PhysRevA.93.033809} and $A$ and $B$ in the Hamiltonian $H$ are not always constant for different $\beta$.
	Indeed, the two EPs correspond to two slightly different sets of scatter designs as predicted by the local basis theory \cite{PhysRevA.101.053842}, rather than the same design predicted by global basis in the two-mode picture.
	Further development is still needed to improve for theoretically describing such WGM cavities with non-Hermitian scattering physical process with multiple weak scatterers, especially near EPs. Nevertheless, our conclusions are further verified with the numerical simulations by solving Maxwell’s equations considering the propagation laws in different media and materials and subsequent experimental demonstrations.
	
	The far-field polarization is correlated to the symmetry of the near-field distribution inside a cavity.
	For the cavity mode with highly symmetric near-field distribution, such as the L3 photonic crystal cavity, the linear far-field polarization can be predicted according to the parities of their in-plane electric field components \cite{Chalcraft:11,Vukovi2002OptimizationOT}.
	However, asymmetric backscattering redistributes the near field and disturbs the symmetry of mode field distribution, which makes the polarization very complex.
	In contrast to the linear polarization in L3 cavity, elliptical polarizations with different polarization angle and polarization degree are obtained for the two modes in the WGM cavities with non-Hermitian mode coupling.
	The numerical simulation results of typical cases are presented in the insets in Fig. \ref{p2}(b).
	To describe the difference between two polarization ellipses, we consider both polarization angle and polarization degree and define
	\begin{equation}
		D_{far}= \frac{\lvert P_1-P_2\rvert}{P_1 + P_2}+\frac{P_1+P_2}{2}\sin{(\Delta \gamma)}.
	\end{equation}
	$P_1$ and $P_2$ are the polarization degree of two modes, defined as the ratio of the major axis $a$ to the minor axis $b$ in the polarization ellipse, while $\Delta \gamma$ is the polarization angle difference (0$^\circ$ to 90$^\circ$) between two modes, which are illustrated in Fig. S4(a) in supplementary materials.
	The first term of $D_{far}$ describes the difference originating from the polarization degree difference between two modes, while the second term describes the difference originating from the polarization angle difference between two modes.
	For the second term, we multiply the polarization angle difference $\sin{(\Delta \gamma)}$ by the averaged polarization degree $(P_1+P_2)/2$.
	This is because when approaching the circular polarization ($P_{1,2} \approx 0$), the difference in the polarization angle has less physical meaning.
	As shown in Fig. \ref{p2}(b), $D_{far}$ decreases when approaching an EP and reaches the minimum value at an EP.
	The polar maps of two polarization ellipses can vary from relatively large $\Delta \gamma$ with $P_1 \ne P_2$ to relatively small $\Delta \gamma$ also with $P_1 \ne P_2$, and coalesce as $\Delta \gamma=0$ and $P_1=P_2$ at the EP.

	\subsection{Experimental results}
		
	\begin{figure*}[htbp]
		\centering
		\includegraphics[scale=0.28]{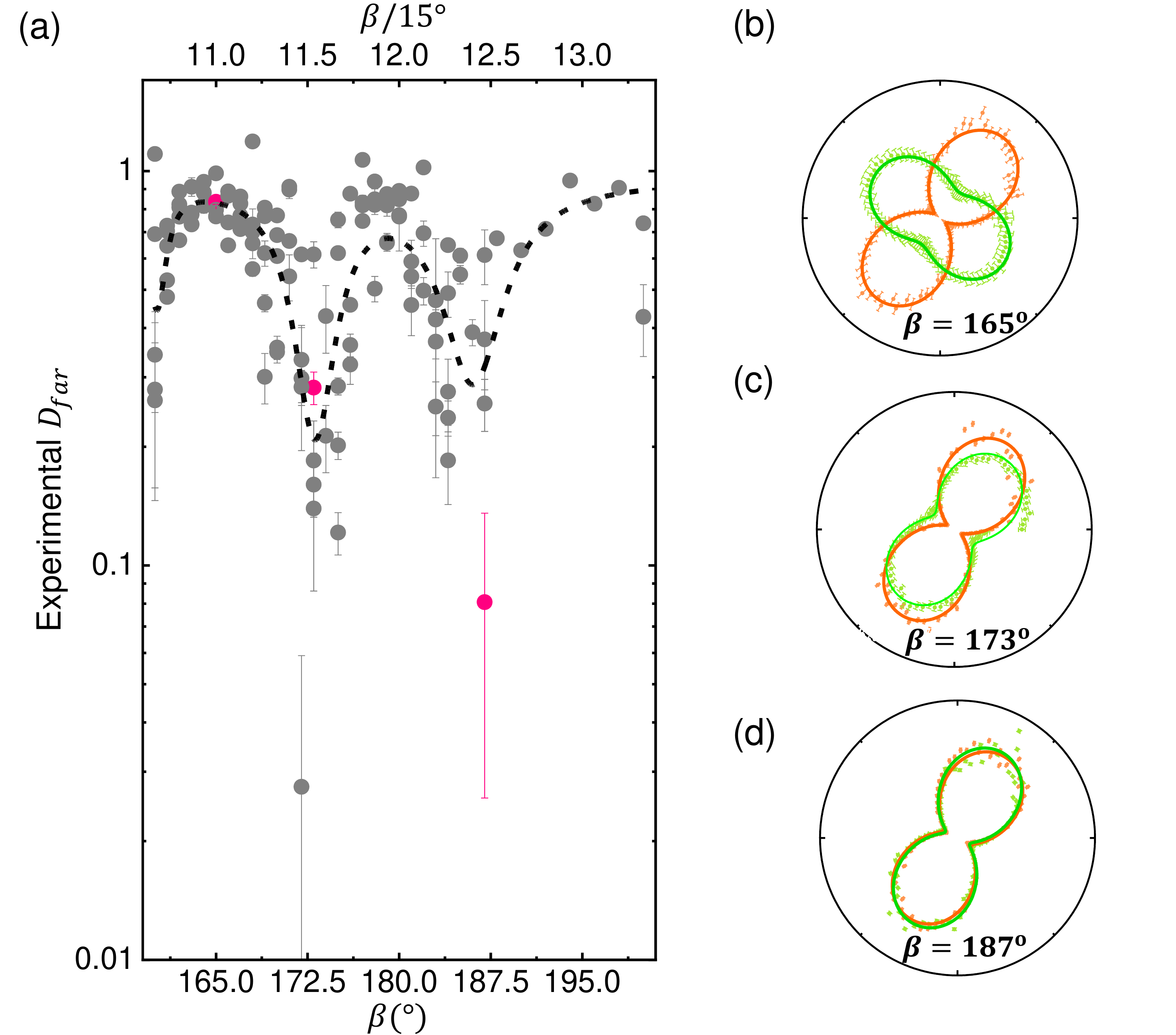}
		\caption{Experimental far-field polarization results from fabricated microcavities with different $\beta$. (a) Experimental $D_{far}$ labeled by gray and pink dots between two polarization ellipses of two modes varying with $\beta$. The dashed line is a fitting curve. (b)-(d) Polar maps of two normalized polarization ellipses for $\beta=165^{\circ}$, $\beta=173^{\circ}$ and $\beta=187^{\circ}$ corresponding to the pink dots in (a). The green and orange dots refer to the experimental PL intensity fitting data with error bars when rotating the $1/2 \lambda$ waveplate and the solid lines refer to theoretical fits. The relative larger error bars in (b) are due to the weaker PL intensities in experiments.}
		\label{p4}
	\end{figure*}

	To experimentally demonstrate the far-field features, we fabricated microcavities with two holes according to the simulation results.
	We span $\beta$ from 160$^{\circ}$ to 200$^{\circ}$ with 1$^{\circ}$ increment, and fabricate five identical microdisks for each parameter.
	Figure \ref{p3}(a) shows the typical SEM images, recorded for two microdisks with different $\beta$.
	We use InAs quantum dots embedded inside the microdisks as the broad-band light sources for the far-field excitation of the cavity modes.
	In principle, all the cavity modes in the spectral range of PL emission from ensemble of quantum dots (QDs) can be observed and be tuned for achieving EPs if with proper scatterer design.
	However, we only focus on $TE_{1,12}$ mode in this work.
	Specific details on the fabrication method and experimental setup can be found in Methods.

	Figure \ref{p3}(b) shows the PL spectra recorded from the microdisks with different $\beta$.
	As shown, we observe that a pair of split peaks gradually emerge into one peak at $\beta=173^{\circ}$ and $\beta=187^{\circ}$, in agreement with the $\beta \approx11.5\times15^{\circ}$ and $\beta \approx12.5\times15^{\circ}$ for realizing EPs predicted in theory discussed above.
	Figure \ref{p3}(c)-(d) shows the statistics of the fitting results of the frequency splitting (c) and the decay rates (d) of the two modes.
	Here in Fig. \ref{p3}(c) we use the difference between two mode frequencies, because the absolute cavity resonance wavelengths contains fluctuations arising from random fabrication errors and intrinsic backscattering \cite{Yang2020}.
	
	In Fig. \ref{p4}(a) we present the experimental results of far-field polarization extracted from the PL spectra.
	As shown, $D_{far}$ varies quasi-periodically with $\beta$ with two clear dips in Fig. \ref{p4}(a), corresponding to the coalescent polar maps of polarization ellipses predicted approaching EPs (Fig. \ref{p2}(b)).
	The polarizations in three typical cases are presented in Fig. \ref{p4}(b)-(d), corresponding to the three pink dots in Fig. \ref{p4}(a).
	The case of $\beta=165^{\circ}=11\times15^{\circ}$ in Fig. \ref{p4}(b) corresponds to a small spatial overlap (165$^\circ$ in Fig. \ref{p2}(a)) with the maximum frequency splitting and decay rate difference.
	As expected, the experimental polarization of spectra (Fig. \ref{p4}(b)) shows that the polarization angle difference can reach up to $\Delta\gamma=\pi/2$, in good agreement with the prediction in Fig. \ref{p2}(b).
	In contrast, for $\beta=173^{\circ}\approx11.5\times15^{\circ}$ in Fig. \ref{p4}(c), polar maps of two polarization ellipses have almost the same polarization angle but different polarization degree.
	For $\beta=187^{\circ}\approx12.5\times15^{\circ}$ in Fig. \ref{p4}(d), polar maps of two polarization ellipses are almost identical.
	These cases are consistent to the corresponding cases approaching EP as predicted in Fig. \ref{p2}(b).
	The good agreement between the theory and the experiment further strengthens our conclusion that the far-field polarization is modulated by the eigenstate non-orthogonality in an optical microcavity with non-Hermitian mode coupling.
	
	\section{Discussion}
	Due to the small scale of holes, fabricated active microdisks with two holes can still allow both small mode volume and high quality factors as shown in supplementary Figure S3(b).
	As they are non-Hermitian system with controllable near-field distribution and far-field polarization, such microcavities show great application potential in polarization-controllable low-threshold lasers or single-photon sources, and in the study of non-linear optics and cavity QED.
	Furthermore, using internal degrees of freedom of light, such as frequency and orbital angular momentum, researchers have proposed to use coupling between different modes to configure synthetic dimensions \cite{Yuan:18}, which can be used for simulating one-dimensional or two-dimensional  non-Hermitian topological systems \cite{PhysRevLett.118.083603, PhysRevLett.122.083903, doi:10.1126/science.aaz3071, PhysRevLett.123.116104}.
	For the cavity in this work, due to the effects of holes on all the modes, $m$ in Eq. \eqref{matrix1} can represent any different mode.
	Therefore, by introducing a coupling between different modes, extra synthetic dimensions could be configured for simulating non-Hermitian topological systems to study rich physical phenomena, such as PT phase transition, non-Hermitian skin effect and novel topological states.
	The far field polarization can provide observable channels for understanding the intrinsic topological physics of these phenomena \cite{10.3389/fphy.2022.862962, PhysRevLett.123.116104}.
	
	In summary, we integrate two different scatterers into active microdisks and obtain controllable non-Hermitian systems with non-orthogonal eigenstates. The eigenstate non-orthogonality is revealed by the spatial field overlap and the far field polarization of cavity modes through numerical simulations where high non-orthogonality corresponds to a large spatial-field overlap and a small difference in polarization angle and polarization degree. Further experimental measurements of far-field polarization of the PL spectra of cavity modes demonstrate such far-field polarization features predicted in simulations. Experimental demonstration of controllable non-Hermitian systems can inspire more integrated optical devices based on controllable non-orthogonal eigenstates. Meanwhile, such microcavities with non-Hermitian mode coupling have potential applications in simulating high-dimensional non-Hermitian topological systems using extra synthetic dimensions to study novel physical phenomena.
	
	\section{Methods}
	\subsection{Theoretical calculations of eigenvalues}
	To analyze how the scatterers affect the microdisks, we calculate the eigenvalues and eigenstates for the two different scatterers. We directly set values to $A$ and $B$, which actually considers the mutual effects on backscattering from each another. $A$ and $B$ are obtained by calculating the eigenvalues at $\beta=165^{\circ}$ and $\beta\approx172.44^{\circ}$ close to an EP. For $\beta=165^{\circ}$, we have $\delta (\omega)=2\sqrt{A^{2}+B^{2}}$. For $\beta_{EP}\approx172.44^{\circ}$ near an EP, we have $\delta (\omega)=2\sqrt{A^{2}+B^{2}+2ABcos(2m)\beta_{EP}}$. By calculating these two equations, we can deduce the complex values of $A$ and $B$. Here $A=0.2985+0.0135i$, $B=0.2990+0.0056i$. While if we use eigenvalues for $\beta_{EP}\approx172.44^{\circ}$ and $\beta\approx187.19^{\circ}$, we have $A=0.0231+0.0089i$ and $B= 0.0234+0.0077i$. Obviously, two sets of $A$ and $B$ are different. For the two different scatterers, $r_{1}=0.037285$ $\mu$m, $r_{2}=0.03203$ $\mu$m, $d_{1}=0.73312$ $\mu$m, $d_{2}=0.74995$ $\mu$m.
	
	\subsection{ Simulation method for spatial field}
	To numerically compute the complex eigenfrequencies and spatial field of $TE_{1,12}$ quasinormal modes in a microdisk, we use a simulation method based on the finite element method to study the wave optics in a whispering galley mode (WGM) microcavity.
	The model we simulate contains a microdisk with two holes, an outer air layer (a column with radius of 2.2 $\mu$m and height of 2.65 $\mu$m removing the microdisk in the center) and a perfectly matched layer (PML) (a column with radius of 2.8 $\mu$m and height of 4.0 $\mu$m removing the microdisk and air layer) which absorbs the waves emitting outwards in different directions and in different wavelength in spectra from the microdisk.
	The refractive index of GaAs for simulation is $n=3.44$.
	Complex eigenfrequencies ($\omega$) of modes can be directly obtained, of which real and imaginary parts represent the resonance frequency and the energy loss rate, respectively.
	The imaginary parts mainly result from the radiation loss due to the curved edge of the microdisk as well as the scattering loss from scatterers.
	The Q factor is calculated by $Q=\frac{Re{(\omega)}}{2Im{(\omega)}}$.

\subsection{Normalization of simulated spatial field}
To calculate the overlap of the spatial field of two quasinormal modes, we need to post-process the simulation results to obtain the integral of two normalized field components from two different quasinormal modes.
The mode field we obtain from the simulation are initially not normalized.
For the normalization of simulation results, we follow the quantized cavity mode field, which is
\begin{equation}
	E(r)=i \sqrt{\frac{\hbar\omega}{2\varepsilon_{r} \varepsilon_{0}}}(a \alpha(r)-a^{+} \alpha^{*}(r)).
\end{equation}
$\omega$ is the cavity mode frequency.
$\varepsilon_{0}$ ($\varepsilon_{r}$) is the vacuum (relative) permittivity.
$a / a^{+}$ is the ladder operator of photons.
$\alpha(r)$ is the normalized cavity mode function which means $\int \alpha (r) \alpha^{*} (r)dV=1$.
Therefore, based on the simulated electric field $E_{cal}$, we normalize and get the cavity mode function as
\begin{equation}
	\alpha(r)= \frac{n(r) E_{cal}(r)}{\sqrt{sum}}.
\end{equation}
Here $sum=\int \lvert n(r) E_{cal}(r) \rvert^{2}dV$.
Then we get the normalized electric field $E(r)$ by
\begin{equation}
	E(r)=\sqrt{\frac{\hbar\omega}{2\varepsilon_{r}\varepsilon_{0}}}\alpha(r).
\end{equation}
Finally we have
\begin{equation}
	E(r)=\sqrt{\frac{\hbar\omega}{2\varepsilon_{0}sum}}E_{cal}(r).
\end{equation}
Here normalization coefficient is $c=\sqrt{\frac{\hbar\omega}{2\varepsilon_{0}sum}}$.

After the normalization of the simulated field, the spatial field overlaps between field components of two quasinormal modes are calculated.
Considering the non-Hermiticity of quasinormal modes, the volume integration to calculate $sum=\int \lvert n(r) E_{cal}(r) \rvert^{2}dV$ and spatial field overlap are both considered PML, according to PML-based normalization in \cite{10.1002/lpor.201700113}.
The refractive index $n(r)$ is set to different values according to the specific material with $n(r)=1$ for air and PML while $n(r)=3.44$ for GaAs.
$V$ includes all the simulation space including the air layer, the PML and the microdisk with two holes.
Therefore, $\lvert \int H_{z1}^{*}H_{z2}dV\rvert$, $\lvert \int E_{x1}^{*}E_{x2}dV\rvert$ and $\lvert \int E_{y1}^{*}E_{y2}dV\rvert$ can be calculated as $S_{H_{z}}$, $S_{E_{x}}$ and $S_{E_{y}}$, respectively.

	\subsection{Simulation method of far-field polarization}
	To calculate the far-field polarization ellipse, we use another method, the finite difference time domain method (FDTD).
	The simulation model is the same as above.
	Randomly positioned magnetic dipoles with random phases are set inside the microdisk cavity as excitation sources to simulate quantum dots embedded inside microdisks.
	Far-field polarization is exhibited through plotting a polarization ellipse with the ellipse polarization angle and polarization degree based on the $s$ and $p$ polarization components with the zero-order diffraction $(0,0)$ considered.
	\subsection{The sample}
	The sample we used for fabricating the microdisks was grown by molecular beam epitaxy. It is constituted from bottom to top by a GaAs substrate, an AlGaAs layer of 1 $\mu$m in thickness and a GaAs slab of  250 nm in thickness. One layer of InAs QDs is grown in the middle of the GaAs slab, which works as excitation sources with a spectral emission range from 250 to 333 THz.
	\subsection{The fabrication process}
	To obtain fabricated microdisks, a mask on the sample surface was first patterned with the designed structures by the electron beam lithography. Then the sample with the mask was etched by inductively coupled plasma to transfer the pattern on the mask onto the GaAs slab and the AlGaAs layer. Finally the sample with the mask removed was etched by HF solutions where most of AlGaAs under the microdisk is etched away. In the end, microdisk devices with an AlGaAs pedestal below were obtained.
	
	\subsection{Optical measurement}
	The optical measurement of fabricated microdisks was implemented with a conventional confocal micro-photoluminescence (PL). The microdisk devices were positioned in a sealed chamber cooled down to 10 K by the liquid helium flow. The QDs were first indirectly excited by a laser with an emission wavelength of 532 nm, and then the PL emission of  the ensemble of QDs in a wide spectral range further excite the microdisk modes. The radiating PL spectra from microdisks were collected by a high-NA objective lens and then reflected and focused by a series of optical elements onto a linear array of InGaAs detectors dispersed through a spectrometer. The far-field polarization of the cavity modes was measured through putting a rotating 1/2$\lambda$ waveplate and a polarizer in the light route and in front of the spectrometer. After traveling through 1/2$\lambda$ waveplate, the polarization rotating angle of PL related to the polarizer, corresponds to twice the rotating angle  $\alpha_{1/2 \lambda}$ of the 1/2$\lambda$ waveplate, which is  $2\alpha_{1/2 \lambda}$.
	
	\subsection{Experimental polarization data analysis}
	The cavity modes exhibits elliptic polarization in the far-field. For a polarization ellipse, polarization degree $P_{i}$ (polarizability) and polarization angle $\gamma_{i}$ (major axis angle) are  used to describe it. Here $P_{i}$ is obtained by fitting the intensity curve of PL spectra of two modes. The data is obtained through multiple-peak Lorentz fitting the intensities of PL spectra of two modes when varying the rotating angle $\alpha_{1/2 \lambda}$ of 1/2$\lambda$ wave-plate,  as shown in Fig. S4 (a) and (b) in supplementary note 4. Then a sine function $y=y0+\lvert A \rvert \rm{sin}(a(\gamma-\gamma_{c}))$ is used to fit the intensity curve with varying $2\alpha_{1/2 \lambda}$. Finally, the polarization degree $P_{i}$ is obtained by calculating $P_{i}=(\lvert A \rvert+y0)/(\lvert A \rvert-y0)$ and the polarization angle difference $\Delta \gamma$ is obtained by calculating $\Delta \gamma=\gamma_{2}-\gamma_{1}$. The major axis is $a=(\lvert A \rvert+y0)/2$ and the minor axis $b=(\lvert A \rvert-y0)/2$. The polarization ellipse is normalized by normalizing the area $S=\pi(\lvert A \rvert+y0)/(\lvert A \rvert-y0)/4$ to 1 as shown in Fig. \ref{p4}. Inversely, the polar maps of polarization ellipses in Fig. \ref{p2}(b) can be obtained when polarization degree and polarization angle are known. To describe the difference between two polarization ellipses, we define a parameter by considering both the differences in major angle and polarization degree as
	\begin{equation}
		D_{far}= \lvert(P_1-P_2) / (P_1 + P_2)\rvert+0.5(P_1+P_2)\sin{(\Delta \gamma)}.
	\end{equation}
	Here $P_1$ and $P_2$ are the polarization degree of two quasinormal modes defined by the ratio of long axis to short axis in the polarization ellipse. $\lvert(P_1-P_2) / (P_1 + P_2)\rvert$ describes the difference in polarization degree for two quasinormal modes and $0.5(P_1+P_2)\sin{(\Delta \gamma)}$ describes the difference in polarization angle multiplying the average polarization degree of two quasinormal modes. For example, for two identical polarization ellipses, $D_{far}= 0$. For two polarization ellipse with different polarization degree and $\Delta \gamma=\pi/2$, $D_{far}= \lvert(P_1-P_2) / (P_1 + P_2)\rvert+0.5(P_1+P_2)$.
	
	\section{Author Contributions}
	X. X., Q. G., K. J. supervised the research project. J. Y. designed and fabricated the devices and conducted the optical measurements. S. S., S. Y., R. Z., B. F. and X. C. helped perform the experiments. , X. X., X. Z., Y. Q., Z. Z.  conducted the theoretical discussion and data analysis. J. Y. and H. L. conducted and discussed the numerical simulation for far-field polarization. All the authors have contributed to writing the manuscript.
	
	\section{Competing interests}
	The authors declare no competing interests.
	
	\section{Data availability}
	Relevant data supporting the key findings of this study are available within the article and the Supplementary Information file. All raw data generated during the current study are available from the corresponding authors upon request.
	\section{\label{sec6}Acknowledgments}
	This work was supported by the National Key Research and Development Program of China (Grant No. 2021YFA1400700), the National Natural Science Foundation of China (Grants Nos. 62025507, 11934019, 92250301, 11721404, 62175254 and 12204020), the Strategic Priority Research Program (Grant No. XDB28000000) of the Chinese Academy of Sciences, China Postdoctoral Science Foundation (Grant No. 2022M710234) and Fundamental Research Funds for the Central Universities, China (Grant No. TP22098A1).
	
\section{References}
%



\begin{thebibliography}{63}%
\makeatletter
\providecommand \@ifxundefined [1]{%
 \@ifx{#1\undefined}
}%
\providecommand \@ifnum [1]{%
 \ifnum #1\expandafter \@firstoftwo
 \else \expandafter \@secondoftwo
 \fi
}%
\providecommand \@ifx [1]{%
 \ifx #1\expandafter \@firstoftwo
 \else \expandafter \@secondoftwo
 \fi
}%
\providecommand \natexlab [1]{#1}%
\providecommand \enquote  [1]{``#1''}%
\providecommand \bibnamefont  [1]{#1}%
\providecommand \bibfnamefont [1]{#1}%
\providecommand \citenamefont [1]{#1}%
\providecommand \href@noop [0]{\@secondoftwo}%
\providecommand \href [0]{\begingroup \@sanitize@url \@href}%
\providecommand \@href[1]{\@@startlink{#1}\@@href}%
\providecommand \@@href[1]{\endgroup#1\@@endlink}%
\providecommand \@sanitize@url [0]{\catcode `\\12\catcode `\$12\catcode
  `\&12\catcode `\#12\catcode `\^12\catcode `\_12\catcode `\%12\relax}%
\providecommand \@@startlink[1]{}%
\providecommand \@@endlink[0]{}%
\providecommand \url  [0]{\begingroup\@sanitize@url \@url }%
\providecommand \@url [1]{\endgroup\@href {#1}{\urlprefix }}%
\providecommand \urlprefix  [0]{URL }%
\providecommand \Eprint [0]{\href }%
\providecommand \doibase [0]{https://doi.org/}%
\providecommand \selectlanguage [0]{\@gobble}%
\providecommand \bibinfo  [0]{\@secondoftwo}%
\providecommand \bibfield  [0]{\@secondoftwo}%
\providecommand \translation [1]{[#1]}%
\providecommand \BibitemOpen [0]{}%
\providecommand \bibitemStop [0]{}%
\providecommand \bibitemNoStop [0]{.\EOS\space}%
\providecommand \EOS [0]{\spacefactor3000\relax}%
\providecommand \BibitemShut  [1]{\csname bibitem#1\endcsname}%
\let\auto@bib@innerbib\@empty
\bibitem [{\citenamefont {Ashida}\ \emph {et~al.}(2020)\citenamefont {Ashida},
  \citenamefont {Gong},\ and\ \citenamefont
  {Ueda}}]{doi:10.1080/00018732.2021.1876991}%
  \BibitemOpen
  \bibfield  {author} {\bibinfo {author} {\bibfnamefont {Y.}~\bibnamefont
  {Ashida}}, \bibinfo {author} {\bibfnamefont {Z.}~\bibnamefont {Gong}},\ and\
  \bibinfo {author} {\bibfnamefont {M.}~\bibnamefont {Ueda}},\ }\bibfield
  {title} {\bibinfo {title} {{Non-Hermitian} physics},\ }\href
  {https://doi.org/10.1080/00018732.2021.1876991} {\bibfield  {journal}
  {\bibinfo  {journal} {Adv. Phys.}\ }\textbf {\bibinfo {volume} {69}},\
  \bibinfo {pages} {249} (\bibinfo {year} {2020})}\BibitemShut {NoStop}%
\bibitem [{\citenamefont {Rotter}(2009)}]{Rotter_2009}%
  \BibitemOpen
  \bibfield  {author} {\bibinfo {author} {\bibfnamefont {I.}~\bibnamefont
  {Rotter}},\ }\bibfield  {title} {\bibinfo {title} {A {non-Hermitian} hamilton
  operator and the physics of open quantum systems},\ }\href
  {https://doi.org/10.1088/1751-8113/42/15/153001} {\bibfield  {journal}
  {\bibinfo  {journal} {J. Phys. A Math. Theor.}\ }\textbf {\bibinfo {volume}
  {42}},\ \bibinfo {pages} {153001} (\bibinfo {year} {2009})}\BibitemShut
  {NoStop}%
\bibitem [{\citenamefont {Gao}\ \emph {et~al.}(2015)\citenamefont {Gao},
  \citenamefont {Estrecho}, \citenamefont {Bliokh}, \citenamefont {Liew},
  \citenamefont {Fraser}, \citenamefont {Brodbeck}, \citenamefont {Kamp},
  \citenamefont {Schneider}, \citenamefont {H{\"o}fling}, \citenamefont
  {Yamamoto}, \citenamefont {Nori}, \citenamefont {Kivshar}, \citenamefont
  {Truscott}, \citenamefont {Dall},\ and\ \citenamefont
  {Ostrovskaya}}]{Gao2015}%
  \BibitemOpen
  \bibfield  {author} {\bibinfo {author} {\bibfnamefont {T.}~\bibnamefont
  {Gao}}, \bibinfo {author} {\bibfnamefont {E.}~\bibnamefont {Estrecho}},
  \bibinfo {author} {\bibfnamefont {K.~Y.}\ \bibnamefont {Bliokh}}, \bibinfo
  {author} {\bibfnamefont {T.~C.~H.}\ \bibnamefont {Liew}}, \bibinfo {author}
  {\bibfnamefont {M.~D.}\ \bibnamefont {Fraser}}, \bibinfo {author}
  {\bibfnamefont {S.}~\bibnamefont {Brodbeck}}, \bibinfo {author}
  {\bibfnamefont {M.}~\bibnamefont {Kamp}}, \bibinfo {author} {\bibfnamefont
  {C.}~\bibnamefont {Schneider}}, \bibinfo {author} {\bibfnamefont
  {S.}~\bibnamefont {H{\"o}fling}}, \bibinfo {author} {\bibfnamefont
  {Y.}~\bibnamefont {Yamamoto}}, \bibinfo {author} {\bibfnamefont
  {F.}~\bibnamefont {Nori}}, \bibinfo {author} {\bibfnamefont {Y.~S.}\
  \bibnamefont {Kivshar}}, \bibinfo {author} {\bibfnamefont {A.~G.}\
  \bibnamefont {Truscott}}, \bibinfo {author} {\bibfnamefont {R.~G.}\
  \bibnamefont {Dall}},\ and\ \bibinfo {author} {\bibfnamefont {E.~A.}\
  \bibnamefont {Ostrovskaya}},\ }\bibfield  {title} {\bibinfo {title}
  {Observation of {non-Hermitian} degeneracies in a chaotic exciton-polariton
  billiard},\ }\href {https://doi.org/10.1038/nature15522} {\bibfield
  {journal} {\bibinfo  {journal} {Nature}\ }\textbf {\bibinfo {volume} {526}},\
  \bibinfo {pages} {554} (\bibinfo {year} {2015})}\BibitemShut {NoStop}%
\bibitem [{\citenamefont {Dembowski}\ \emph {et~al.}(2004)\citenamefont
  {Dembowski}, \citenamefont {Dietz}, \citenamefont {Gr\"af}, \citenamefont
  {Harney}, \citenamefont {Heine}, \citenamefont {Heiss},\ and\ \citenamefont
  {Richter}}]{PhysRevE.69.056216}%
  \BibitemOpen
  \bibfield  {author} {\bibinfo {author} {\bibfnamefont {C.}~\bibnamefont
  {Dembowski}}, \bibinfo {author} {\bibfnamefont {B.}~\bibnamefont {Dietz}},
  \bibinfo {author} {\bibfnamefont {H.-D.}\ \bibnamefont {Gr\"af}}, \bibinfo
  {author} {\bibfnamefont {H.~L.}\ \bibnamefont {Harney}}, \bibinfo {author}
  {\bibfnamefont {A.}~\bibnamefont {Heine}}, \bibinfo {author} {\bibfnamefont
  {W.~D.}\ \bibnamefont {Heiss}},\ and\ \bibinfo {author} {\bibfnamefont
  {A.}~\bibnamefont {Richter}},\ }\bibfield  {title} {\bibinfo {title}
  {Encircling an exceptional point},\ }\href
  {https://doi.org/10.1103/PhysRevE.69.056216} {\bibfield  {journal} {\bibinfo
  {journal} {Phys. Rev. E}\ }\textbf {\bibinfo {volume} {69}},\ \bibinfo
  {pages} {056216} (\bibinfo {year} {2004})}\BibitemShut {NoStop}%
\bibitem [{\citenamefont {Cartarius}\ \emph {et~al.}(2007)\citenamefont
  {Cartarius}, \citenamefont {Main},\ and\ \citenamefont
  {Wunner}}]{PhysRevLett.99.173003}%
  \BibitemOpen
  \bibfield  {author} {\bibinfo {author} {\bibfnamefont {H.}~\bibnamefont
  {Cartarius}}, \bibinfo {author} {\bibfnamefont {J.}~\bibnamefont {Main}},\
  and\ \bibinfo {author} {\bibfnamefont {G.}~\bibnamefont {Wunner}},\
  }\bibfield  {title} {\bibinfo {title} {Exceptional points in atomic
  spectra},\ }\href {https://doi.org/10.1103/PhysRevLett.99.173003} {\bibfield
  {journal} {\bibinfo  {journal} {Phys. Rev. Lett.}\ }\textbf {\bibinfo
  {volume} {99}},\ \bibinfo {pages} {173003} (\bibinfo {year}
  {2007})}\BibitemShut {NoStop}%
\bibitem [{\citenamefont {Ding}\ \emph {et~al.}(2018)\citenamefont {Ding},
  \citenamefont {Ma}, \citenamefont {Zhang},\ and\ \citenamefont
  {Chan}}]{PhysRevLett.121.085702}%
  \BibitemOpen
  \bibfield  {author} {\bibinfo {author} {\bibfnamefont {K.}~\bibnamefont
  {Ding}}, \bibinfo {author} {\bibfnamefont {G.}~\bibnamefont {Ma}}, \bibinfo
  {author} {\bibfnamefont {Z.~Q.}\ \bibnamefont {Zhang}},\ and\ \bibinfo
  {author} {\bibfnamefont {C.~T.}\ \bibnamefont {Chan}},\ }\bibfield  {title}
  {\bibinfo {title} {Experimental demonstration of an anisotropic exceptional
  point},\ }\href {https://doi.org/10.1103/PhysRevLett.121.085702} {\bibfield
  {journal} {\bibinfo  {journal} {Phys. Rev. Lett.}\ }\textbf {\bibinfo
  {volume} {121}},\ \bibinfo {pages} {085702} (\bibinfo {year}
  {2018})}\BibitemShut {NoStop}%
\bibitem [{\citenamefont {Soleymani}\ \emph {et~al.}(2022)\citenamefont
  {Soleymani}, \citenamefont {Zhong}, \citenamefont {Mokim}, \citenamefont
  {Rotter}, \citenamefont {El-Ganainy},\ and\ \citenamefont
  {{\"O}zdemir}}]{Soleymani2022}%
  \BibitemOpen
  \bibfield  {author} {\bibinfo {author} {\bibfnamefont {S.}~\bibnamefont
  {Soleymani}}, \bibinfo {author} {\bibfnamefont {Q.}~\bibnamefont {Zhong}},
  \bibinfo {author} {\bibfnamefont {M.}~\bibnamefont {Mokim}}, \bibinfo
  {author} {\bibfnamefont {S.}~\bibnamefont {Rotter}}, \bibinfo {author}
  {\bibfnamefont {R.}~\bibnamefont {El-Ganainy}},\ and\ \bibinfo {author}
  {\bibfnamefont {{\c{S}}.~K.}\ \bibnamefont {{\"O}zdemir}},\ }\bibfield
  {title} {\bibinfo {title} {Chiral and degenerate perfect absorption on
  exceptional surfaces},\ }\href {https://doi.org/10.1038/s41467-022-27990-w}
  {\bibfield  {journal} {\bibinfo  {journal} {Nat. Commun.}\ }\textbf {\bibinfo
  {volume} {13}},\ \bibinfo {pages} {599} (\bibinfo {year} {2022})}\BibitemShut
  {NoStop}%
\bibitem [{\citenamefont {Wang}\ \emph {et~al.}(2020)\citenamefont {Wang},
  \citenamefont {Jiang}, \citenamefont {Zhao}, \citenamefont {Zhang},
  \citenamefont {Hsu}, \citenamefont {Peng}, \citenamefont {Stone},
  \citenamefont {Jiang},\ and\ \citenamefont {Yang}}]{Wang2020}%
  \BibitemOpen
  \bibfield  {author} {\bibinfo {author} {\bibfnamefont {C.}~\bibnamefont
  {Wang}}, \bibinfo {author} {\bibfnamefont {X.}~\bibnamefont {Jiang}},
  \bibinfo {author} {\bibfnamefont {G.}~\bibnamefont {Zhao}}, \bibinfo {author}
  {\bibfnamefont {M.}~\bibnamefont {Zhang}}, \bibinfo {author} {\bibfnamefont
  {C.~W.}\ \bibnamefont {Hsu}}, \bibinfo {author} {\bibfnamefont
  {B.}~\bibnamefont {Peng}}, \bibinfo {author} {\bibfnamefont {A.~D.}\
  \bibnamefont {Stone}}, \bibinfo {author} {\bibfnamefont {L.}~\bibnamefont
  {Jiang}},\ and\ \bibinfo {author} {\bibfnamefont {L.}~\bibnamefont {Yang}},\
  }\bibfield  {title} {\bibinfo {title} {Electromagnetically induced
  transparency at a chiral exceptional point},\ }\href
  {https://doi.org/10.1038/s41567-019-0746-7} {\bibfield  {journal} {\bibinfo
  {journal} {Nat. Phys.}\ }\textbf {\bibinfo {volume} {16}},\ \bibinfo {pages}
  {334} (\bibinfo {year} {2020})}\BibitemShut {NoStop}%
\bibitem [{\citenamefont {Lu}\ \emph {et~al.}(2022)\citenamefont {Lu},
  \citenamefont {Zhao}, \citenamefont {Li},\ and\ \citenamefont {Liu}}]{Lu:22}%
  \BibitemOpen
  \bibfield  {author} {\bibinfo {author} {\bibfnamefont {Y.}~\bibnamefont
  {Lu}}, \bibinfo {author} {\bibfnamefont {Y.}~\bibnamefont {Zhao}}, \bibinfo
  {author} {\bibfnamefont {R.}~\bibnamefont {Li}},\ and\ \bibinfo {author}
  {\bibfnamefont {J.}~\bibnamefont {Liu}},\ }\bibfield  {title} {\bibinfo
  {title} {Anomalous spontaneous emission dynamics at chiral exceptional
  points},\ }\href {https://doi.org/10.1364/OE.473824} {\bibfield  {journal}
  {\bibinfo  {journal} {Opt. Express}\ }\textbf {\bibinfo {volume} {30}},\
  \bibinfo {pages} {41784} (\bibinfo {year} {2022})}\BibitemShut {NoStop}%
\bibitem [{\citenamefont {Kim}\ \emph {et~al.}(2021)\citenamefont {Kim},
  \citenamefont {Kim}, \citenamefont {Seo}, \citenamefont {Park}, \citenamefont
  {Moon},\ and\ \citenamefont {An}}]{Kim2021}%
  \BibitemOpen
  \bibfield  {author} {\bibinfo {author} {\bibfnamefont {J.}~\bibnamefont
  {Kim}}, \bibinfo {author} {\bibfnamefont {J.}~\bibnamefont {Kim}}, \bibinfo
  {author} {\bibfnamefont {J.}~\bibnamefont {Seo}}, \bibinfo {author}
  {\bibfnamefont {K.-W.}\ \bibnamefont {Park}}, \bibinfo {author}
  {\bibfnamefont {S.}~\bibnamefont {Moon}},\ and\ \bibinfo {author}
  {\bibfnamefont {K.}~\bibnamefont {An}},\ }\bibfield  {title} {\bibinfo
  {title} {Practical lineshape of a laser operating near an exceptional
  point},\ }\href {https://doi.org/10.1038/s41598-021-85665-w} {\bibfield
  {journal} {\bibinfo  {journal} {Sci. Rep.}\ }\textbf {\bibinfo {volume}
  {11}},\ \bibinfo {pages} {6164} (\bibinfo {year} {2021})}\BibitemShut
  {NoStop}%
\bibitem [{\citenamefont {Peng}\ \emph {et~al.}(2016)\citenamefont {Peng},
  \citenamefont {{\"O}zdemir}, \citenamefont {Liertzer}, \citenamefont {Chen},
  \citenamefont {Kramer}, \citenamefont {Yılmaz}, \citenamefont {Wiersig},
  \citenamefont {Rotter},\ and\ \citenamefont
  {Yang}}]{doi:10.1073/pnas.1603318113}%
  \BibitemOpen
  \bibfield  {author} {\bibinfo {author} {\bibfnamefont {B.}~\bibnamefont
  {Peng}}, \bibinfo {author} {\bibfnamefont {{\c S}.~K.}\ \bibnamefont
  {{\"O}zdemir}}, \bibinfo {author} {\bibfnamefont {M.}~\bibnamefont
  {Liertzer}}, \bibinfo {author} {\bibfnamefont {W.}~\bibnamefont {Chen}},
  \bibinfo {author} {\bibfnamefont {J.}~\bibnamefont {Kramer}}, \bibinfo
  {author} {\bibfnamefont {H.}~\bibnamefont {Yılmaz}}, \bibinfo {author}
  {\bibfnamefont {J.}~\bibnamefont {Wiersig}}, \bibinfo {author} {\bibfnamefont
  {S.}~\bibnamefont {Rotter}},\ and\ \bibinfo {author} {\bibfnamefont
  {L.}~\bibnamefont {Yang}},\ }\bibfield  {title} {\bibinfo {title} {Chiral
  modes and directional lasing at exceptional points},\ }\href
  {https://doi.org/10.1073/pnas.1603318113} {\bibfield  {journal} {\bibinfo
  {journal} {Proc. Natl. Acad. Sci. U.S.A.}\ }\textbf {\bibinfo {volume}
  {113}},\ \bibinfo {pages} {6845} (\bibinfo {year} {2016})}\BibitemShut
  {NoStop}%
\bibitem [{\citenamefont {Chen}\ \emph {et~al.}(2022)\citenamefont {Chen},
  \citenamefont {Zhang}, \citenamefont {Zhao}, \citenamefont {Wu},
  \citenamefont {Fang}, \citenamefont {Yang},\ and\ \citenamefont
  {Nori}}]{PhysRevA.106.022438}%
  \BibitemOpen
  \bibfield  {author} {\bibinfo {author} {\bibfnamefont {D.-X.}\ \bibnamefont
  {Chen}}, \bibinfo {author} {\bibfnamefont {Y.}~\bibnamefont {Zhang}},
  \bibinfo {author} {\bibfnamefont {J.-L.}\ \bibnamefont {Zhao}}, \bibinfo
  {author} {\bibfnamefont {Q.-C.}\ \bibnamefont {Wu}}, \bibinfo {author}
  {\bibfnamefont {Y.-L.}\ \bibnamefont {Fang}}, \bibinfo {author}
  {\bibfnamefont {C.-P.}\ \bibnamefont {Yang}},\ and\ \bibinfo {author}
  {\bibfnamefont {F.}~\bibnamefont {Nori}},\ }\bibfield  {title} {\bibinfo
  {title} {Quantum state discrimination in a $\mathcal{PT}$-symmetric system},\
  }\href {https://doi.org/10.1103/PhysRevA.106.022438} {\bibfield  {journal}
  {\bibinfo  {journal} {Phys. Rev. A}\ }\textbf {\bibinfo {volume} {106}},\
  \bibinfo {pages} {022438} (\bibinfo {year} {2022})}\BibitemShut {NoStop}%
\bibitem [{\citenamefont {idemann}\ \emph {et~al.}(2020)\citenamefont
  {idemann}, \citenamefont {Kremer}, \citenamefont {Helbig}, \citenamefont
  {Hofmann}, \citenamefont {Stegmaier}, \citenamefont {Greiter}, \citenamefont
  {Thomale},\ and\ \citenamefont {Szameit}}]{doi:10.1126/science.aaz8727}%
  \BibitemOpen
  \bibfield  {author} {\bibinfo {author} {\bibfnamefont {S.}~\bibnamefont
  {idemann}}, \bibinfo {author} {\bibfnamefont {M.}~\bibnamefont {Kremer}},
  \bibinfo {author} {\bibfnamefont {T.}~\bibnamefont {Helbig}}, \bibinfo
  {author} {\bibfnamefont {T.}~\bibnamefont {Hofmann}}, \bibinfo {author}
  {\bibfnamefont {A.}~\bibnamefont {Stegmaier}}, \bibinfo {author}
  {\bibfnamefont {M.}~\bibnamefont {Greiter}}, \bibinfo {author} {\bibfnamefont
  {R.}~\bibnamefont {Thomale}},\ and\ \bibinfo {author} {\bibfnamefont
  {A.}~\bibnamefont {Szameit}},\ }\bibfield  {title} {\bibinfo {title}
  {Topological funneling of light},\ }\href
  {https://doi.org/10.1126/science.aaz8727} {\bibfield  {journal} {\bibinfo
  {journal} {Science}\ }\textbf {\bibinfo {volume} {368}},\ \bibinfo {pages}
  {311} (\bibinfo {year} {2020})}\BibitemShut {NoStop}%
\bibitem [{\citenamefont {Ding}\ \emph {et~al.}(2016)\citenamefont {Ding},
  \citenamefont {Ma}, \citenamefont {Xiao}, \citenamefont {Zhang},\ and\
  \citenamefont {Chan}}]{PhysRevX.6.021007}%
  \BibitemOpen
  \bibfield  {author} {\bibinfo {author} {\bibfnamefont {K.}~\bibnamefont
  {Ding}}, \bibinfo {author} {\bibfnamefont {G.}~\bibnamefont {Ma}}, \bibinfo
  {author} {\bibfnamefont {M.}~\bibnamefont {Xiao}}, \bibinfo {author}
  {\bibfnamefont {Z.~Q.}\ \bibnamefont {Zhang}},\ and\ \bibinfo {author}
  {\bibfnamefont {C.~T.}\ \bibnamefont {Chan}},\ }\bibfield  {title} {\bibinfo
  {title} {Emergence, coalescence, and topological properties of multiple
  exceptional points and their experimental realization},\ }\href
  {https://doi.org/10.1103/PhysRevX.6.021007} {\bibfield  {journal} {\bibinfo
  {journal} {Phys. Rev. X}\ }\textbf {\bibinfo {volume} {6}},\ \bibinfo {pages}
  {021007} (\bibinfo {year} {2016})}\BibitemShut {NoStop}%
\bibitem [{\citenamefont {Miri}\ and\ \citenamefont
  {Alù}(2019)}]{doi:10.1126/science.aar7709}%
  \BibitemOpen
  \bibfield  {author} {\bibinfo {author} {\bibfnamefont {M.-A.}\ \bibnamefont
  {Miri}}\ and\ \bibinfo {author} {\bibfnamefont {A.}~\bibnamefont {Alù}},\
  }\bibfield  {title} {\bibinfo {title} {Exceptional points in optics and
  photonics},\ }\href {https://doi.org/10.1126/science.aar7709} {\bibfield
  {journal} {\bibinfo  {journal} {Science}\ }\textbf {\bibinfo {volume}
  {363}},\ \bibinfo {pages} {eaar7709} (\bibinfo {year} {2019})}\BibitemShut
  {NoStop}%
\bibitem [{\citenamefont {Feng}\ \emph {et~al.}(2017)\citenamefont {Feng},
  \citenamefont {El-Ganainy},\ and\ \citenamefont {Ge}}]{Feng2017}%
  \BibitemOpen
  \bibfield  {author} {\bibinfo {author} {\bibfnamefont {L.}~\bibnamefont
  {Feng}}, \bibinfo {author} {\bibfnamefont {R.}~\bibnamefont {El-Ganainy}},\
  and\ \bibinfo {author} {\bibfnamefont {L.}~\bibnamefont {Ge}},\ }\bibfield
  {title} {\bibinfo {title} {{Non-Hermitian} photonics based on parity--time
  symmetry},\ }\href {https://doi.org/10.1038/s41566-017-0031-1} {\bibfield
  {journal} {\bibinfo  {journal} {Nat. Photonics}\ }\textbf {\bibinfo {volume}
  {11}},\ \bibinfo {pages} {752} (\bibinfo {year} {2017})}\BibitemShut
  {NoStop}%
\bibitem [{\citenamefont {Hassan}\ \emph {et~al.}(2017)\citenamefont {Hassan},
  \citenamefont {Zhen}, \citenamefont {Solja\ifmmode \check{c}\else
  \v{c}\fi{}i\ifmmode~\acute{c}\else \'{c}\fi{}}, \citenamefont {Khajavikhan},\
  and\ \citenamefont {Christodoulides}}]{PhysRevLett.118.093002}%
  \BibitemOpen
  \bibfield  {author} {\bibinfo {author} {\bibfnamefont {A.~U.}\ \bibnamefont
  {Hassan}}, \bibinfo {author} {\bibfnamefont {B.}~\bibnamefont {Zhen}},
  \bibinfo {author} {\bibfnamefont {M.}~\bibnamefont {Solja\ifmmode
  \check{c}\else \v{c}\fi{}i\ifmmode~\acute{c}\else \'{c}\fi{}}}, \bibinfo
  {author} {\bibfnamefont {M.}~\bibnamefont {Khajavikhan}},\ and\ \bibinfo
  {author} {\bibfnamefont {D.~N.}\ \bibnamefont {Christodoulides}},\ }\bibfield
   {title} {\bibinfo {title} {Dynamically encircling exceptional points : Exact
  evolution and polarization state conversion},\ }\href
  {https://doi.org/10.1103/PhysRevLett.118.093002} {\bibfield  {journal}
  {\bibinfo  {journal} {Phys. Rev. Lett.}\ }\textbf {\bibinfo {volume} {118}},\
  \bibinfo {pages} {093002} (\bibinfo {year} {2017})}\BibitemShut {NoStop}%
\bibitem [{\citenamefont {Hassani~Gangaraj}\ and\ \citenamefont
  {Monticone}(2018)}]{PhysRevLett.121.093901}%
  \BibitemOpen
  \bibfield  {author} {\bibinfo {author} {\bibfnamefont {S.~A.}\ \bibnamefont
  {Hassani~Gangaraj}}\ and\ \bibinfo {author} {\bibfnamefont {F.}~\bibnamefont
  {Monticone}},\ }\bibfield  {title} {\bibinfo {title} {Topological waveguiding
  near an exceptional point : Defect-immune, slow-light, and loss-immune
  propagation},\ }\href {https://doi.org/10.1103/PhysRevLett.121.093901}
  {\bibfield  {journal} {\bibinfo  {journal} {Phys. Rev. Lett.}\ }\textbf
  {\bibinfo {volume} {121}},\ \bibinfo {pages} {093901} (\bibinfo {year}
  {2018})}\BibitemShut {NoStop}%
\bibitem [{\citenamefont {Li}\ \emph {et~al.}(2022)\citenamefont {Li},
  \citenamefont {Liang}, \citenamefont {Wang}, \citenamefont {Lu},\ and\
  \citenamefont {Liu}}]{PhysRevLett.128.223903}%
  \BibitemOpen
  \bibfield  {author} {\bibinfo {author} {\bibfnamefont {Y.}~\bibnamefont
  {Li}}, \bibinfo {author} {\bibfnamefont {C.}~\bibnamefont {Liang}}, \bibinfo
  {author} {\bibfnamefont {C.}~\bibnamefont {Wang}}, \bibinfo {author}
  {\bibfnamefont {C.}~\bibnamefont {Lu}},\ and\ \bibinfo {author}
  {\bibfnamefont {Y.-C.}\ \bibnamefont {Liu}},\ }\bibfield  {title} {\bibinfo
  {title} {Gain-loss-induced hybrid skin-topological effect},\ }\href
  {https://doi.org/10.1103/PhysRevLett.128.223903} {\bibfield  {journal}
  {\bibinfo  {journal} {Phys. Rev. Lett.}\ }\textbf {\bibinfo {volume} {128}},\
  \bibinfo {pages} {223903} (\bibinfo {year} {2022})}\BibitemShut {NoStop}%
\bibitem [{\citenamefont {Wiersig}(2011)}]{PhysRevA.84.063828}%
  \BibitemOpen
  \bibfield  {author} {\bibinfo {author} {\bibfnamefont {J.}~\bibnamefont
  {Wiersig}},\ }\bibfield  {title} {\bibinfo {title} {Structure of
  whispering-gallery modes in optical microdisks perturbed by nanoparticles},\
  }\href {https://doi.org/10.1103/PhysRevA.84.063828} {\bibfield  {journal}
  {\bibinfo  {journal} {Phys. Rev. A}\ }\textbf {\bibinfo {volume} {84}},\
  \bibinfo {pages} {063828} (\bibinfo {year} {2011})}\BibitemShut {NoStop}%
\bibitem [{\citenamefont {Zhang}\ \emph {et~al.}(2020)\citenamefont {Zhang},
  \citenamefont {Feng}, \citenamefont {Chen}, \citenamefont {Ge},\ and\
  \citenamefont {Wan}}]{PhysRevLett.124.053901}%
  \BibitemOpen
  \bibfield  {author} {\bibinfo {author} {\bibfnamefont {F.}~\bibnamefont
  {Zhang}}, \bibinfo {author} {\bibfnamefont {Y.}~\bibnamefont {Feng}},
  \bibinfo {author} {\bibfnamefont {X.}~\bibnamefont {Chen}}, \bibinfo {author}
  {\bibfnamefont {L.}~\bibnamefont {Ge}},\ and\ \bibinfo {author}
  {\bibfnamefont {W.}~\bibnamefont {Wan}},\ }\bibfield  {title} {\bibinfo
  {title} {Synthetic anti-{PT} symmetry in a single microcavity},\ }\href
  {https://doi.org/10.1103/PhysRevLett.124.053901} {\bibfield  {journal}
  {\bibinfo  {journal} {Phys. Rev. Lett.}\ }\textbf {\bibinfo {volume} {124}},\
  \bibinfo {pages} {053901} (\bibinfo {year} {2020})}\BibitemShut {NoStop}%
\bibitem [{\citenamefont {Biasi}\ \emph {et~al.}(2022)\citenamefont {Biasi},
  \citenamefont {Franchi}, \citenamefont {Mione},\ and\ \citenamefont
  {Pavesi}}]{Biasi:22}%
  \BibitemOpen
  \bibfield  {author} {\bibinfo {author} {\bibfnamefont {S.}~\bibnamefont
  {Biasi}}, \bibinfo {author} {\bibfnamefont {R.}~\bibnamefont {Franchi}},
  \bibinfo {author} {\bibfnamefont {F.}~\bibnamefont {Mione}},\ and\ \bibinfo
  {author} {\bibfnamefont {L.}~\bibnamefont {Pavesi}},\ }\bibfield  {title}
  {\bibinfo {title} {Interferometric method to estimate the eigenvalues of a
  non-hermitian two-level optical system},\ }\href
  {https://doi.org/10.1364/PRJ.450402} {\bibfield  {journal} {\bibinfo
  {journal} {Photon. Res.}\ }\textbf {\bibinfo {volume} {10}},\ \bibinfo
  {pages} {1134} (\bibinfo {year} {2022})}\BibitemShut {NoStop}%
\bibitem [{\citenamefont {Chen}\ \emph {et~al.}(2017)\citenamefont {Chen},
  \citenamefont {Kaya~{\"O}zdemir}, \citenamefont {Zhao}, \citenamefont
  {Wiersig},\ and\ \citenamefont {Yang}}]{Chen2017}%
  \BibitemOpen
  \bibfield  {author} {\bibinfo {author} {\bibfnamefont {W.}~\bibnamefont
  {Chen}}, \bibinfo {author} {\bibfnamefont {{\c{S}}.}~\bibnamefont
  {Kaya~{\"O}zdemir}}, \bibinfo {author} {\bibfnamefont {G.}~\bibnamefont
  {Zhao}}, \bibinfo {author} {\bibfnamefont {J.}~\bibnamefont {Wiersig}},\ and\
  \bibinfo {author} {\bibfnamefont {L.}~\bibnamefont {Yang}},\ }\bibfield
  {title} {\bibinfo {title} {Exceptional points enhance sensing in an optical
  microcavity},\ }\href {https://doi.org/10.1038/nature23281} {\bibfield
  {journal} {\bibinfo  {journal} {Nature}\ }\textbf {\bibinfo {volume} {548}},\
  \bibinfo {pages} {192} (\bibinfo {year} {2017})}\BibitemShut {NoStop}%
\bibitem [{\citenamefont {Makris}\ \emph {et~al.}(2008)\citenamefont {Makris},
  \citenamefont {El-Ganainy}, \citenamefont {Christodoulides},\ and\
  \citenamefont {Musslimani}}]{PhysRevLett.100.103904}%
  \BibitemOpen
  \bibfield  {author} {\bibinfo {author} {\bibfnamefont {K.~G.}\ \bibnamefont
  {Makris}}, \bibinfo {author} {\bibfnamefont {R.}~\bibnamefont {El-Ganainy}},
  \bibinfo {author} {\bibfnamefont {D.~N.}\ \bibnamefont {Christodoulides}},\
  and\ \bibinfo {author} {\bibfnamefont {Z.~H.}\ \bibnamefont {Musslimani}},\
  }\bibfield  {title} {\bibinfo {title} {Beam dynamics in
  $\mathcal{P}\mathcal{T}$ symmetric optical lattices},\ }\href
  {https://doi.org/10.1103/PhysRevLett.100.103904} {\bibfield  {journal}
  {\bibinfo  {journal} {Phys. Rev. Lett.}\ }\textbf {\bibinfo {volume} {100}},\
  \bibinfo {pages} {103904} (\bibinfo {year} {2008})}\BibitemShut {NoStop}%
\bibitem [{\citenamefont {Petermann}(1979)}]{1070064}%
  \BibitemOpen
  \bibfield  {author} {\bibinfo {author} {\bibfnamefont {K.}~\bibnamefont
  {Petermann}},\ }\bibfield  {title} {\bibinfo {title} {Calculated spontaneous
  emission factor for double-heterostructure injection lasers with gain-induced
  waveguiding},\ }\href {https://doi.org/10.1109/JQE.1979.1070064} {\bibfield
  {journal} {\bibinfo  {journal} {IEEE J. Quantum Electron.}\ }\textbf
  {\bibinfo {volume} {15}},\ \bibinfo {pages} {566} (\bibinfo {year}
  {1979})}\BibitemShut {NoStop}%
\bibitem [{\citenamefont {Vuckovic}\ \emph {et~al.}(2002)\citenamefont
  {Vuckovic}, \citenamefont {Loncar}, \citenamefont {Mabuchi},\ and\
  \citenamefont {Scherer}}]{Vukovi2002OptimizationOT}%
  \BibitemOpen
  \bibfield  {author} {\bibinfo {author} {\bibfnamefont {J.}~\bibnamefont
  {Vuckovic}}, \bibinfo {author} {\bibfnamefont {M.}~\bibnamefont {Loncar}},
  \bibinfo {author} {\bibfnamefont {H.}~\bibnamefont {Mabuchi}},\ and\ \bibinfo
  {author} {\bibfnamefont {A.}~\bibnamefont {Scherer}},\ }\bibfield  {title}
  {\bibinfo {title} {Optimization of the {Q} factor in photonic crystal
  microcavities},\ }\href@noop {} {\bibfield  {journal} {\bibinfo  {journal}
  {IEEE J. Quantum Electron.}\ }\textbf {\bibinfo {volume} {38}},\ \bibinfo
  {pages} {850} (\bibinfo {year} {2002})}\BibitemShut {NoStop}%
\bibitem [{\citenamefont {Chalcraft}\ \emph {et~al.}(2011)\citenamefont
  {Chalcraft}, \citenamefont {Lam}, \citenamefont {Jones}, \citenamefont
  {Szymanski}, \citenamefont {Oulton}, \citenamefont {Thijssen}, \citenamefont
  {Skolnick}, \citenamefont {Whittaker}, \citenamefont {Krauss},\ and\
  \citenamefont {Fox}}]{Chalcraft:11}%
  \BibitemOpen
  \bibfield  {author} {\bibinfo {author} {\bibfnamefont {A.~R.~A.}\
  \bibnamefont {Chalcraft}}, \bibinfo {author} {\bibfnamefont {S.}~\bibnamefont
  {Lam}}, \bibinfo {author} {\bibfnamefont {B.~D.}\ \bibnamefont {Jones}},
  \bibinfo {author} {\bibfnamefont {D.}~\bibnamefont {Szymanski}}, \bibinfo
  {author} {\bibfnamefont {R.}~\bibnamefont {Oulton}}, \bibinfo {author}
  {\bibfnamefont {A.~C.~T.}\ \bibnamefont {Thijssen}}, \bibinfo {author}
  {\bibfnamefont {M.~S.}\ \bibnamefont {Skolnick}}, \bibinfo {author}
  {\bibfnamefont {D.~M.}\ \bibnamefont {Whittaker}}, \bibinfo {author}
  {\bibfnamefont {T.~F.}\ \bibnamefont {Krauss}},\ and\ \bibinfo {author}
  {\bibfnamefont {A.~M.}\ \bibnamefont {Fox}},\ }\bibfield  {title} {\bibinfo
  {title} {Mode structure of coupled {L3} photonic crystal cavities},\ }\href
  {https://doi.org/10.1364/OE.19.005670} {\bibfield  {journal} {\bibinfo
  {journal} {Opt. Express}\ }\textbf {\bibinfo {volume} {19}},\ \bibinfo
  {pages} {5670} (\bibinfo {year} {2011})}\BibitemShut {NoStop}%
\bibitem [{\citenamefont {Xiong}\ \emph {et~al.}(2018)\citenamefont {Xiong},
  \citenamefont {Hsu}, \citenamefont {Bromberg}, \citenamefont {Antonio-Lopez},
  \citenamefont {Amezcua~Correa},\ and\ \citenamefont {Cao}}]{Xiong2018}%
  \BibitemOpen
  \bibfield  {author} {\bibinfo {author} {\bibfnamefont {W.}~\bibnamefont
  {Xiong}}, \bibinfo {author} {\bibfnamefont {C.~W.}\ \bibnamefont {Hsu}},
  \bibinfo {author} {\bibfnamefont {Y.}~\bibnamefont {Bromberg}}, \bibinfo
  {author} {\bibfnamefont {J.~E.}\ \bibnamefont {Antonio-Lopez}}, \bibinfo
  {author} {\bibfnamefont {R.}~\bibnamefont {Amezcua~Correa}},\ and\ \bibinfo
  {author} {\bibfnamefont {H.}~\bibnamefont {Cao}},\ }\bibfield  {title}
  {\bibinfo {title} {Complete polarization control in multimode fibers with
  polarization and mode coupling},\ }\href
  {https://doi.org/10.1038/s41377-018-0047-4} {\bibfield  {journal} {\bibinfo
  {journal} {Light: Science {\&} Applications}\ }\textbf {\bibinfo {volume}
  {7}},\ \bibinfo {pages} {54} (\bibinfo {year} {2018})}\BibitemShut {NoStop}%
\bibitem [{\citenamefont {Wang}\ \emph {et~al.}(2019)\citenamefont {Wang},
  \citenamefont {He}, \citenamefont {Chung}, \citenamefont {Hu}, \citenamefont
  {Yu}, \citenamefont {Chen}, \citenamefont {Ding}, \citenamefont {Chen},
  \citenamefont {Qin}, \citenamefont {Yang}, \citenamefont {Liu}, \citenamefont
  {Duan}, \citenamefont {Li}, \citenamefont {Gerhardt}, \citenamefont
  {Winkler}, \citenamefont {Jurkat}, \citenamefont {Wang}, \citenamefont
  {Gregersen}, \citenamefont {Huo}, \citenamefont {Dai}, \citenamefont {Yu},
  \citenamefont {H{\"o}fling}, \citenamefont {Lu},\ and\ \citenamefont
  {Pan}}]{Wang2019}%
  \BibitemOpen
  \bibfield  {author} {\bibinfo {author} {\bibfnamefont {H.}~\bibnamefont
  {Wang}}, \bibinfo {author} {\bibfnamefont {Y.-M.}\ \bibnamefont {He}},
  \bibinfo {author} {\bibfnamefont {T.-H.}\ \bibnamefont {Chung}}, \bibinfo
  {author} {\bibfnamefont {H.}~\bibnamefont {Hu}}, \bibinfo {author}
  {\bibfnamefont {Y.}~\bibnamefont {Yu}}, \bibinfo {author} {\bibfnamefont
  {S.}~\bibnamefont {Chen}}, \bibinfo {author} {\bibfnamefont {X.}~\bibnamefont
  {Ding}}, \bibinfo {author} {\bibfnamefont {M.-C.}\ \bibnamefont {Chen}},
  \bibinfo {author} {\bibfnamefont {J.}~\bibnamefont {Qin}}, \bibinfo {author}
  {\bibfnamefont {X.}~\bibnamefont {Yang}}, \bibinfo {author} {\bibfnamefont
  {R.-Z.}\ \bibnamefont {Liu}}, \bibinfo {author} {\bibfnamefont {Z.-C.}\
  \bibnamefont {Duan}}, \bibinfo {author} {\bibfnamefont {J.-P.}\ \bibnamefont
  {Li}}, \bibinfo {author} {\bibfnamefont {S.}~\bibnamefont {Gerhardt}},
  \bibinfo {author} {\bibfnamefont {K.}~\bibnamefont {Winkler}}, \bibinfo
  {author} {\bibfnamefont {J.}~\bibnamefont {Jurkat}}, \bibinfo {author}
  {\bibfnamefont {L.-J.}\ \bibnamefont {Wang}}, \bibinfo {author}
  {\bibfnamefont {N.}~\bibnamefont {Gregersen}}, \bibinfo {author}
  {\bibfnamefont {Y.-H.}\ \bibnamefont {Huo}}, \bibinfo {author} {\bibfnamefont
  {Q.}~\bibnamefont {Dai}}, \bibinfo {author} {\bibfnamefont {S.}~\bibnamefont
  {Yu}}, \bibinfo {author} {\bibfnamefont {S.}~\bibnamefont {H{\"o}fling}},
  \bibinfo {author} {\bibfnamefont {C.-Y.}\ \bibnamefont {Lu}},\ and\ \bibinfo
  {author} {\bibfnamefont {J.-W.}\ \bibnamefont {Pan}},\ }\bibfield  {title}
  {\bibinfo {title} {Towards optimal single-photon sources from polarized
  microcavities},\ }\href {https://doi.org/10.1038/s41566-019-0494-3}
  {\bibfield  {journal} {\bibinfo  {journal} {Nat. Photonics}\ }\textbf
  {\bibinfo {volume} {13}},\ \bibinfo {pages} {770} (\bibinfo {year}
  {2019})}\BibitemShut {NoStop}%
\bibitem [{\citenamefont {He}\ \emph {et~al.}(2016)\citenamefont {He},
  \citenamefont {Ma}, \citenamefont {Cui}, \citenamefont {Yu}, \citenamefont
  {Yang}, \citenamefont {Song}, \citenamefont {Wu}, \citenamefont {Chen},
  \citenamefont {Chen},\ and\ \citenamefont {Qian}}]{He2016}%
  \BibitemOpen
  \bibfield  {author} {\bibinfo {author} {\bibfnamefont {H.}~\bibnamefont
  {He}}, \bibinfo {author} {\bibfnamefont {E.}~\bibnamefont {Ma}}, \bibinfo
  {author} {\bibfnamefont {Y.}~\bibnamefont {Cui}}, \bibinfo {author}
  {\bibfnamefont {J.}~\bibnamefont {Yu}}, \bibinfo {author} {\bibfnamefont
  {Y.}~\bibnamefont {Yang}}, \bibinfo {author} {\bibfnamefont {T.}~\bibnamefont
  {Song}}, \bibinfo {author} {\bibfnamefont {C.-D.}\ \bibnamefont {Wu}},
  \bibinfo {author} {\bibfnamefont {X.}~\bibnamefont {Chen}}, \bibinfo {author}
  {\bibfnamefont {B.}~\bibnamefont {Chen}},\ and\ \bibinfo {author}
  {\bibfnamefont {G.}~\bibnamefont {Qian}},\ }\bibfield  {title} {\bibinfo
  {title} {Polarized three-photon-pumped laser in a single {MOF}
  microcrystal},\ }\href {https://doi.org/10.1038/ncomms11087} {\bibfield
  {journal} {\bibinfo  {journal} {Nat. Commun.}\ }\textbf {\bibinfo {volume}
  {7}},\ \bibinfo {pages} {11087} (\bibinfo {year} {2016})}\BibitemShut
  {NoStop}%
\bibitem [{\citenamefont {Weissflog}\ \emph {et~al.}(2022)\citenamefont
  {Weissflog}, \citenamefont {Cai}, \citenamefont {Parry}, \citenamefont
  {Rahmani}, \citenamefont {Xu}, \citenamefont {Arslan}, \citenamefont
  {Fedotova}, \citenamefont {Marino}, \citenamefont {Lysevych}, \citenamefont
  {Tan}, \citenamefont {Jagadish}, \citenamefont {Miroshnichenko},
  \citenamefont {Leo}, \citenamefont {Sukhorukov}, \citenamefont {Setzpfandt},
  \citenamefont {Pertsch}, \citenamefont {Staude},\ and\ \citenamefont
  {Neshev}}]{10.1002/lpor.202200183}%
  \BibitemOpen
  \bibfield  {author} {\bibinfo {author} {\bibfnamefont {M.~A.}\ \bibnamefont
  {Weissflog}}, \bibinfo {author} {\bibfnamefont {M.}~\bibnamefont {Cai}},
  \bibinfo {author} {\bibfnamefont {M.}~\bibnamefont {Parry}}, \bibinfo
  {author} {\bibfnamefont {M.}~\bibnamefont {Rahmani}}, \bibinfo {author}
  {\bibfnamefont {L.}~\bibnamefont {Xu}}, \bibinfo {author} {\bibfnamefont
  {D.}~\bibnamefont {Arslan}}, \bibinfo {author} {\bibfnamefont
  {A.}~\bibnamefont {Fedotova}}, \bibinfo {author} {\bibfnamefont
  {G.}~\bibnamefont {Marino}}, \bibinfo {author} {\bibfnamefont
  {M.}~\bibnamefont {Lysevych}}, \bibinfo {author} {\bibfnamefont {H.~H.}\
  \bibnamefont {Tan}}, \bibinfo {author} {\bibfnamefont {C.}~\bibnamefont
  {Jagadish}}, \bibinfo {author} {\bibfnamefont {A.}~\bibnamefont
  {Miroshnichenko}}, \bibinfo {author} {\bibfnamefont {G.}~\bibnamefont {Leo}},
  \bibinfo {author} {\bibfnamefont {A.~A.}\ \bibnamefont {Sukhorukov}},
  \bibinfo {author} {\bibfnamefont {F.}~\bibnamefont {Setzpfandt}}, \bibinfo
  {author} {\bibfnamefont {T.}~\bibnamefont {Pertsch}}, \bibinfo {author}
  {\bibfnamefont {I.}~\bibnamefont {Staude}},\ and\ \bibinfo {author}
  {\bibfnamefont {D.~N.}\ \bibnamefont {Neshev}},\ }\bibfield  {title}
  {\bibinfo {title} {Far-field polarization engineering from nonlinear
  nanoresonators},\ }\href
  {https://doi.org/https://doi.org/10.1002/lpor.202200183} {\bibfield
  {journal} {\bibinfo  {journal} {Laser Photonics Rev.}\ }\textbf {\bibinfo
  {volume} {16}},\ \bibinfo {pages} {2200183} (\bibinfo {year}
  {2022})}\BibitemShut {NoStop}%
\bibitem [{\citenamefont {Wang}\ \emph {et~al.}(2022)\citenamefont {Wang},
  \citenamefont {Yin}, \citenamefont {Zhang}, \citenamefont {Chen},
  \citenamefont {Wang}, \citenamefont {Li}, \citenamefont {Hu}, \citenamefont
  {Zhou},\ and\ \citenamefont {Peng}}]{10.3389/fphy.2022.862962}%
  \BibitemOpen
  \bibfield  {author} {\bibinfo {author} {\bibfnamefont {F.}~\bibnamefont
  {Wang}}, \bibinfo {author} {\bibfnamefont {X.}~\bibnamefont {Yin}}, \bibinfo
  {author} {\bibfnamefont {Z.}~\bibnamefont {Zhang}}, \bibinfo {author}
  {\bibfnamefont {Z.}~\bibnamefont {Chen}}, \bibinfo {author} {\bibfnamefont
  {H.}~\bibnamefont {Wang}}, \bibinfo {author} {\bibfnamefont {P.}~\bibnamefont
  {Li}}, \bibinfo {author} {\bibfnamefont {Y.}~\bibnamefont {Hu}}, \bibinfo
  {author} {\bibfnamefont {X.}~\bibnamefont {Zhou}},\ and\ \bibinfo {author}
  {\bibfnamefont {C.}~\bibnamefont {Peng}},\ }\bibfield  {title} {\bibinfo
  {title} {Fundamentals and applications of topological polarization
  singularities},\ }\bibfield  {journal} {\bibinfo  {journal} {Front. Phys.}\
  }\textbf {\bibinfo {volume} {10}},\ \href
  {https://doi.org/10.3389/fphy.2022.862962} {10.3389/fphy.2022.862962}
  (\bibinfo {year} {2022})\BibitemShut {NoStop}%
\bibitem [{\citenamefont {Forbes}\ \emph {et~al.}(2021)\citenamefont {Forbes},
  \citenamefont {de~Oliveira},\ and\ \citenamefont {Dennis}}]{Forbes2021}%
  \BibitemOpen
  \bibfield  {author} {\bibinfo {author} {\bibfnamefont {A.}~\bibnamefont
  {Forbes}}, \bibinfo {author} {\bibfnamefont {M.}~\bibnamefont
  {de~Oliveira}},\ and\ \bibinfo {author} {\bibfnamefont {M.~R.}\ \bibnamefont
  {Dennis}},\ }\bibfield  {title} {\bibinfo {title} {Structured light},\ }\href
  {https://doi.org/10.1038/s41566-021-00780-4} {\bibfield  {journal} {\bibinfo
  {journal} {Nature Photonics}\ }\textbf {\bibinfo {volume} {15}},\ \bibinfo
  {pages} {253} (\bibinfo {year} {2021})}\BibitemShut {NoStop}%
\bibitem [{\citenamefont {Feng}\ \emph {et~al.}(2014)\citenamefont {Feng},
  \citenamefont {Wong}, \citenamefont {Ma}, \citenamefont {Wang},\ and\
  \citenamefont {Zhang}}]{doi:10.1126/science.1258479}%
  \BibitemOpen
  \bibfield  {author} {\bibinfo {author} {\bibfnamefont {L.}~\bibnamefont
  {Feng}}, \bibinfo {author} {\bibfnamefont {Z.~J.}\ \bibnamefont {Wong}},
  \bibinfo {author} {\bibfnamefont {R.-M.}\ \bibnamefont {Ma}}, \bibinfo
  {author} {\bibfnamefont {Y.}~\bibnamefont {Wang}},\ and\ \bibinfo {author}
  {\bibfnamefont {X.}~\bibnamefont {Zhang}},\ }\bibfield  {title} {\bibinfo
  {title} {Single-mode laser by parity-time symmetry breaking},\ }\href
  {https://doi.org/10.1126/science.1258479} {\bibfield  {journal} {\bibinfo
  {journal} {Science}\ }\textbf {\bibinfo {volume} {346}},\ \bibinfo {pages}
  {972} (\bibinfo {year} {2014})}\BibitemShut {NoStop}%
\bibitem [{\citenamefont {Kim}\ \emph {et~al.}(2014)\citenamefont {Kim},
  \citenamefont {Kwon}, \citenamefont {Shim}, \citenamefont {Jung},\ and\
  \citenamefont {Yu}}]{Kim:14}%
  \BibitemOpen
  \bibfield  {author} {\bibinfo {author} {\bibfnamefont {M.}~\bibnamefont
  {Kim}}, \bibinfo {author} {\bibfnamefont {K.}~\bibnamefont {Kwon}}, \bibinfo
  {author} {\bibfnamefont {J.}~\bibnamefont {Shim}}, \bibinfo {author}
  {\bibfnamefont {Y.}~\bibnamefont {Jung}},\ and\ \bibinfo {author}
  {\bibfnamefont {K.}~\bibnamefont {Yu}},\ }\bibfield  {title} {\bibinfo
  {title} {Partially directional microdisk laser with two {Rayleigh}
  scatterers},\ }\href {https://doi.org/10.1364/OL.39.002423} {\bibfield
  {journal} {\bibinfo  {journal} {Opt. Lett.}\ }\textbf {\bibinfo {volume}
  {39}},\ \bibinfo {pages} {2423} (\bibinfo {year} {2014})}\BibitemShut
  {NoStop}%
\bibitem [{\citenamefont {Wiersig}(2016)}]{PhysRevA.93.033809}%
  \BibitemOpen
  \bibfield  {author} {\bibinfo {author} {\bibfnamefont {J.}~\bibnamefont
  {Wiersig}},\ }\bibfield  {title} {\bibinfo {title} {Sensors operating at
  exceptional points: General theory},\ }\href
  {https://doi.org/10.1103/PhysRevA.93.033809} {\bibfield  {journal} {\bibinfo
  {journal} {Phys. Rev. A}\ }\textbf {\bibinfo {volume} {93}},\ \bibinfo
  {pages} {033809} (\bibinfo {year} {2016})}\BibitemShut {NoStop}%
\bibitem [{\citenamefont {Yang}\ \emph {et~al.}(2021)\citenamefont {Yang},
  \citenamefont {Shi}, \citenamefont {Xie}, \citenamefont {Wu}, \citenamefont
  {Xiao}, \citenamefont {Song}, \citenamefont {Dang}, \citenamefont {Sun},
  \citenamefont {Yang}, \citenamefont {wang}, \citenamefont {Ge}, \citenamefont
  {Li}, \citenamefont {Zuo}, \citenamefont {Jin},\ and\ \citenamefont
  {Xu}}]{Yang:21}%
  \BibitemOpen
  \bibfield  {author} {\bibinfo {author} {\bibfnamefont {J.}~\bibnamefont
  {Yang}}, \bibinfo {author} {\bibfnamefont {S.}~\bibnamefont {Shi}}, \bibinfo
  {author} {\bibfnamefont {X.}~\bibnamefont {Xie}}, \bibinfo {author}
  {\bibfnamefont {S.}~\bibnamefont {Wu}}, \bibinfo {author} {\bibfnamefont
  {S.}~\bibnamefont {Xiao}}, \bibinfo {author} {\bibfnamefont {F.}~\bibnamefont
  {Song}}, \bibinfo {author} {\bibfnamefont {J.}~\bibnamefont {Dang}}, \bibinfo
  {author} {\bibfnamefont {S.}~\bibnamefont {Sun}}, \bibinfo {author}
  {\bibfnamefont {L.}~\bibnamefont {Yang}}, \bibinfo {author} {\bibfnamefont
  {Y.}~\bibnamefont {wang}}, \bibinfo {author} {\bibfnamefont {Z.-Y.}\
  \bibnamefont {Ge}}, \bibinfo {author} {\bibfnamefont {B.-B.}\ \bibnamefont
  {Li}}, \bibinfo {author} {\bibfnamefont {Z.}~\bibnamefont {Zuo}}, \bibinfo
  {author} {\bibfnamefont {K.}~\bibnamefont {Jin}},\ and\ \bibinfo {author}
  {\bibfnamefont {X.}~\bibnamefont {Xu}},\ }\bibfield  {title} {\bibinfo
  {title} {Enhanced emission from a single quantum dot in a microdisk at a
  deterministic diabolical point},\ }\href {https://doi.org/10.1364/OE.419740}
  {\bibfield  {journal} {\bibinfo  {journal} {Opt. Express}\ }\textbf {\bibinfo
  {volume} {29}},\ \bibinfo {pages} {14231} (\bibinfo {year}
  {2021})}\BibitemShut {NoStop}%
\bibitem [{\citenamefont {Zhu}\ \emph {et~al.}(2023)\citenamefont {Zhu},
  \citenamefont {Wang}, \citenamefont {Tao}, \citenamefont {Fu}, \citenamefont
  {Liu}, \citenamefont {Bo}, \citenamefont {Yang}, \citenamefont {Zhang},\ and\
  \citenamefont {Xu}}]{PhysRevA.108.L041501}%
  \BibitemOpen
  \bibfield  {author} {\bibinfo {author} {\bibfnamefont {J.}~\bibnamefont
  {Zhu}}, \bibinfo {author} {\bibfnamefont {C.}~\bibnamefont {Wang}}, \bibinfo
  {author} {\bibfnamefont {C.}~\bibnamefont {Tao}}, \bibinfo {author}
  {\bibfnamefont {Z.}~\bibnamefont {Fu}}, \bibinfo {author} {\bibfnamefont
  {H.}~\bibnamefont {Liu}}, \bibinfo {author} {\bibfnamefont {F.}~\bibnamefont
  {Bo}}, \bibinfo {author} {\bibfnamefont {L.}~\bibnamefont {Yang}}, \bibinfo
  {author} {\bibfnamefont {G.}~\bibnamefont {Zhang}},\ and\ \bibinfo {author}
  {\bibfnamefont {J.}~\bibnamefont {Xu}},\ }\bibfield  {title} {\bibinfo
  {title} {Local chirality at exceptional points in optical whispering-gallery
  microcavities},\ }\href {https://doi.org/10.1103/PhysRevA.108.L041501}
  {\bibfield  {journal} {\bibinfo  {journal} {Phys. Rev. A}\ }\textbf {\bibinfo
  {volume} {108}},\ \bibinfo {pages} {L041501} (\bibinfo {year}
  {2023})}\BibitemShut {NoStop}%
\bibitem [{\citenamefont {Fong}\ \emph {et~al.}(2021)\citenamefont {Fong},
  \citenamefont {Ota}, \citenamefont {Arakawa}, \citenamefont {Iwamoto},\ and\
  \citenamefont {Kato}}]{PhysRevResearch.3.043096}%
  \BibitemOpen
  \bibfield  {author} {\bibinfo {author} {\bibfnamefont {C.~F.}\ \bibnamefont
  {Fong}}, \bibinfo {author} {\bibfnamefont {Y.}~\bibnamefont {Ota}}, \bibinfo
  {author} {\bibfnamefont {Y.}~\bibnamefont {Arakawa}}, \bibinfo {author}
  {\bibfnamefont {S.}~\bibnamefont {Iwamoto}},\ and\ \bibinfo {author}
  {\bibfnamefont {Y.~K.}\ \bibnamefont {Kato}},\ }\bibfield  {title} {\bibinfo
  {title} {Chiral modes near exceptional points in symmetry broken {H1}
  photonic crystal cavities},\ }\href
  {https://doi.org/10.1103/PhysRevResearch.3.043096} {\bibfield  {journal}
  {\bibinfo  {journal} {Phys. Rev. Res.}\ }\textbf {\bibinfo {volume} {3}},\
  \bibinfo {pages} {043096} (\bibinfo {year} {2021})}\BibitemShut {NoStop}%
\bibitem [{\citenamefont {Khanbekyan}\ and\ \citenamefont
  {Wiersig}(2020)}]{PhysRevResearch.2.023375}%
  \BibitemOpen
  \bibfield  {author} {\bibinfo {author} {\bibfnamefont {M.}~\bibnamefont
  {Khanbekyan}}\ and\ \bibinfo {author} {\bibfnamefont {J.}~\bibnamefont
  {Wiersig}},\ }\bibfield  {title} {\bibinfo {title} {Decay suppression of
  spontaneous emission of a single emitter in a {high-Q} cavity at exceptional
  points},\ }\href {https://doi.org/10.1103/PhysRevResearch.2.023375}
  {\bibfield  {journal} {\bibinfo  {journal} {Phys. Rev. Research}\ }\textbf
  {\bibinfo {volume} {2}},\ \bibinfo {pages} {023375} (\bibinfo {year}
  {2020})}\BibitemShut {NoStop}%
\bibitem [{\citenamefont {Pick}\ \emph {et~al.}(2017)\citenamefont {Pick},
  \citenamefont {Zhen}, \citenamefont {Miller}, \citenamefont {Hsu},
  \citenamefont {Hernandez}, \citenamefont {Rodriguez}, \citenamefont
  {Solja\v{c}i\'{c}},\ and\ \citenamefont {Johnson}}]{Pick:17}%
  \BibitemOpen
  \bibfield  {author} {\bibinfo {author} {\bibfnamefont {A.}~\bibnamefont
  {Pick}}, \bibinfo {author} {\bibfnamefont {B.}~\bibnamefont {Zhen}}, \bibinfo
  {author} {\bibfnamefont {O.~D.}\ \bibnamefont {Miller}}, \bibinfo {author}
  {\bibfnamefont {C.~W.}\ \bibnamefont {Hsu}}, \bibinfo {author} {\bibfnamefont
  {F.}~\bibnamefont {Hernandez}}, \bibinfo {author} {\bibfnamefont {A.~W.}\
  \bibnamefont {Rodriguez}}, \bibinfo {author} {\bibfnamefont {M.}~\bibnamefont
  {Solja\v{c}i\'{c}}},\ and\ \bibinfo {author} {\bibfnamefont {S.~G.}\
  \bibnamefont {Johnson}},\ }\bibfield  {title} {\bibinfo {title} {General
  theory of spontaneous emission near exceptional points},\ }\href
  {https://doi.org/10.1364/OE.25.012325} {\bibfield  {journal} {\bibinfo
  {journal} {Opt. Express}\ }\textbf {\bibinfo {volume} {25}},\ \bibinfo
  {pages} {12325} (\bibinfo {year} {2017})}\BibitemShut {NoStop}%
\bibitem [{\citenamefont {Lin}\ \emph {et~al.}(2016)\citenamefont {Lin},
  \citenamefont {Pick}, \citenamefont {Lon\ifmmode~\check{c}\else
  \v{c}\fi{}ar},\ and\ \citenamefont {Rodriguez}}]{PhysRevLett.117.107402}%
  \BibitemOpen
  \bibfield  {author} {\bibinfo {author} {\bibfnamefont {Z.}~\bibnamefont
  {Lin}}, \bibinfo {author} {\bibfnamefont {A.}~\bibnamefont {Pick}}, \bibinfo
  {author} {\bibfnamefont {M.}~\bibnamefont {Lon\ifmmode~\check{c}\else
  \v{c}\fi{}ar}},\ and\ \bibinfo {author} {\bibfnamefont {A.~W.}\ \bibnamefont
  {Rodriguez}},\ }\bibfield  {title} {\bibinfo {title} {Enhanced spontaneous
  emission at third-order dirac exceptional points in inverse-designed photonic
  crystals},\ }\href {https://doi.org/10.1103/PhysRevLett.117.107402}
  {\bibfield  {journal} {\bibinfo  {journal} {Phys. Rev. Lett.}\ }\textbf
  {\bibinfo {volume} {117}},\ \bibinfo {pages} {107402} (\bibinfo {year}
  {2016})}\BibitemShut {NoStop}%
\bibitem [{\citenamefont {Silberstein}\ \emph {et~al.}(2020)\citenamefont
  {Silberstein}, \citenamefont {Behrends}, \citenamefont {Goldstein},\ and\
  \citenamefont {Ilan}}]{PhysRevB.102.245147}%
  \BibitemOpen
  \bibfield  {author} {\bibinfo {author} {\bibfnamefont {N.}~\bibnamefont
  {Silberstein}}, \bibinfo {author} {\bibfnamefont {J.}~\bibnamefont
  {Behrends}}, \bibinfo {author} {\bibfnamefont {M.}~\bibnamefont
  {Goldstein}},\ and\ \bibinfo {author} {\bibfnamefont {R.}~\bibnamefont
  {Ilan}},\ }\bibfield  {title} {\bibinfo {title} {Berry connection induced
  anomalous wave-packet dynamics in non-hermitian systems},\ }\href
  {https://doi.org/10.1103/PhysRevB.102.245147} {\bibfield  {journal} {\bibinfo
   {journal} {Phys. Rev. B}\ }\textbf {\bibinfo {volume} {102}},\ \bibinfo
  {pages} {245147} (\bibinfo {year} {2020})}\BibitemShut {NoStop}%
\bibitem [{\citenamefont {Bergholtz}\ \emph {et~al.}(2021)\citenamefont
  {Bergholtz}, \citenamefont {Budich},\ and\ \citenamefont
  {Kunst}}]{RevModPhys.93.015005}%
  \BibitemOpen
  \bibfield  {author} {\bibinfo {author} {\bibfnamefont {E.~J.}\ \bibnamefont
  {Bergholtz}}, \bibinfo {author} {\bibfnamefont {J.~C.}\ \bibnamefont
  {Budich}},\ and\ \bibinfo {author} {\bibfnamefont {F.~K.}\ \bibnamefont
  {Kunst}},\ }\bibfield  {title} {\bibinfo {title} {Exceptional topology of
  non-hermitian systems},\ }\href
  {https://doi.org/10.1103/RevModPhys.93.015005} {\bibfield  {journal}
  {\bibinfo  {journal} {Rev. Mod. Phys.}\ }\textbf {\bibinfo {volume} {93}},\
  \bibinfo {pages} {015005} (\bibinfo {year} {2021})}\BibitemShut {NoStop}%
\bibitem [{\citenamefont {Lalanne}\ \emph {et~al.}(2018)\citenamefont
  {Lalanne}, \citenamefont {Yan}, \citenamefont {Vynck}, \citenamefont
  {Sauvan},\ and\ \citenamefont {Hugonin}}]{10.1002/lpor.201700113}%
  \BibitemOpen
  \bibfield  {author} {\bibinfo {author} {\bibfnamefont {P.}~\bibnamefont
  {Lalanne}}, \bibinfo {author} {\bibfnamefont {W.}~\bibnamefont {Yan}},
  \bibinfo {author} {\bibfnamefont {K.}~\bibnamefont {Vynck}}, \bibinfo
  {author} {\bibfnamefont {C.}~\bibnamefont {Sauvan}},\ and\ \bibinfo {author}
  {\bibfnamefont {J.-P.}\ \bibnamefont {Hugonin}},\ }\bibfield  {title}
  {\bibinfo {title} {Light interaction with photonic and plasmonic
  resonances},\ }\href {https://doi.org/https://doi.org/10.1002/lpor.201700113}
  {\bibfield  {journal} {\bibinfo  {journal} {Laser Photonics Rev.}\ }\textbf
  {\bibinfo {volume} {12}},\ \bibinfo {pages} {1700113} (\bibinfo {year}
  {2018})}\BibitemShut {NoStop}%
\bibitem [{\citenamefont {Gigli}\ \emph {et~al.}(2020)\citenamefont {Gigli},
  \citenamefont {Wu}, \citenamefont {Marino}, \citenamefont {Borne},
  \citenamefont {Leo},\ and\ \citenamefont
  {Lalanne}}]{doi:10.1021/acsphotonics.0c00014}%
  \BibitemOpen
  \bibfield  {author} {\bibinfo {author} {\bibfnamefont {C.}~\bibnamefont
  {Gigli}}, \bibinfo {author} {\bibfnamefont {T.}~\bibnamefont {Wu}}, \bibinfo
  {author} {\bibfnamefont {G.}~\bibnamefont {Marino}}, \bibinfo {author}
  {\bibfnamefont {A.}~\bibnamefont {Borne}}, \bibinfo {author} {\bibfnamefont
  {G.}~\bibnamefont {Leo}},\ and\ \bibinfo {author} {\bibfnamefont
  {P.}~\bibnamefont {Lalanne}},\ }\bibfield  {title} {\bibinfo {title}
  {Quasinormal-mode non-hermitian modeling and design in nonlinear
  nano-optics},\ }\href {https://doi.org/10.1021/acsphotonics.0c00014}
  {\bibfield  {journal} {\bibinfo  {journal} {ACS Photonics}\ }\textbf
  {\bibinfo {volume} {7}},\ \bibinfo {pages} {1197} (\bibinfo {year}
  {2020})}\BibitemShut {NoStop}%
\bibitem [{\citenamefont {Lalanne}\ \emph {et~al.}(2019)\citenamefont
  {Lalanne}, \citenamefont {Yan}, \citenamefont {Gras}, \citenamefont {Sauvan},
  \citenamefont {Hugonin}, \citenamefont {Besbes}, \citenamefont {Dem\'{e}sy},
  \citenamefont {Truong}, \citenamefont {Gralak}, \citenamefont {Zolla},
  \citenamefont {Nicolet}, \citenamefont {Binkowski}, \citenamefont
  {Zschiedrich}, \citenamefont {Burger}, \citenamefont {Zimmerling},
  \citenamefont {Remis}, \citenamefont {Urbach}, \citenamefont {Liu},\ and\
  \citenamefont {Weiss}}]{Lalanne:19}%
  \BibitemOpen
  \bibfield  {author} {\bibinfo {author} {\bibfnamefont {P.}~\bibnamefont
  {Lalanne}}, \bibinfo {author} {\bibfnamefont {W.}~\bibnamefont {Yan}},
  \bibinfo {author} {\bibfnamefont {A.}~\bibnamefont {Gras}}, \bibinfo {author}
  {\bibfnamefont {C.}~\bibnamefont {Sauvan}}, \bibinfo {author} {\bibfnamefont
  {J.-P.}\ \bibnamefont {Hugonin}}, \bibinfo {author} {\bibfnamefont
  {M.}~\bibnamefont {Besbes}}, \bibinfo {author} {\bibfnamefont
  {G.}~\bibnamefont {Dem\'{e}sy}}, \bibinfo {author} {\bibfnamefont {M.~D.}\
  \bibnamefont {Truong}}, \bibinfo {author} {\bibfnamefont {B.}~\bibnamefont
  {Gralak}}, \bibinfo {author} {\bibfnamefont {F.}~\bibnamefont {Zolla}},
  \bibinfo {author} {\bibfnamefont {A.}~\bibnamefont {Nicolet}}, \bibinfo
  {author} {\bibfnamefont {F.}~\bibnamefont {Binkowski}}, \bibinfo {author}
  {\bibfnamefont {L.}~\bibnamefont {Zschiedrich}}, \bibinfo {author}
  {\bibfnamefont {S.}~\bibnamefont {Burger}}, \bibinfo {author} {\bibfnamefont
  {J.}~\bibnamefont {Zimmerling}}, \bibinfo {author} {\bibfnamefont
  {R.}~\bibnamefont {Remis}}, \bibinfo {author} {\bibfnamefont
  {P.}~\bibnamefont {Urbach}}, \bibinfo {author} {\bibfnamefont {H.~T.}\
  \bibnamefont {Liu}},\ and\ \bibinfo {author} {\bibfnamefont {T.}~\bibnamefont
  {Weiss}},\ }\bibfield  {title} {\bibinfo {title} {Quasinormal mode solvers
  for resonators with dispersive materials},\ }\href
  {https://doi.org/10.1364/JOSAA.36.000686} {\bibfield  {journal} {\bibinfo
  {journal} {J. Opt. Soc. Am. A}\ }\textbf {\bibinfo {volume} {36}},\ \bibinfo
  {pages} {686} (\bibinfo {year} {2019})}\BibitemShut {NoStop}%
\bibitem [{\citenamefont {Srinivasan}\ \emph {et~al.}(2006)\citenamefont
  {Srinivasan}, \citenamefont {Borselli}, \citenamefont {Painter},
  \citenamefont {Stintz},\ and\ \citenamefont {Krishna}}]{Srinivasan:06}%
  \BibitemOpen
  \bibfield  {author} {\bibinfo {author} {\bibfnamefont {K.}~\bibnamefont
  {Srinivasan}}, \bibinfo {author} {\bibfnamefont {M.}~\bibnamefont
  {Borselli}}, \bibinfo {author} {\bibfnamefont {O.}~\bibnamefont {Painter}},
  \bibinfo {author} {\bibfnamefont {A.}~\bibnamefont {Stintz}},\ and\ \bibinfo
  {author} {\bibfnamefont {S.}~\bibnamefont {Krishna}},\ }\bibfield  {title}
  {\bibinfo {title} {Cavity q, mode volume, and lasing threshold in small
  diameter algaas microdisks with embedded quantum dots},\ }\href
  {https://doi.org/10.1364/OE.14.001094} {\bibfield  {journal} {\bibinfo
  {journal} {Opt. Express}\ }\textbf {\bibinfo {volume} {14}},\ \bibinfo
  {pages} {1094} (\bibinfo {year} {2006})}\BibitemShut {NoStop}%
\bibitem [{\citenamefont {Silva}\ \emph {et~al.}(2008)\citenamefont {Silva},
  \citenamefont {Parra-Murillo}, \citenamefont {Valentim}, \citenamefont
  {Morais}, \citenamefont {Plentz}, \citenamefont {{a}es}, \citenamefont
  {Vinck-Posada}, \citenamefont {Rodriguez}, \citenamefont {Skolnick},
  \citenamefont {Tahraoui},\ and\ \citenamefont {Hopkinson}}]{Silva:08}%
  \BibitemOpen
  \bibfield  {author} {\bibinfo {author} {\bibfnamefont {A.~G.}\ \bibnamefont
  {Silva}}, \bibinfo {author} {\bibfnamefont {C.~A.}\ \bibnamefont
  {Parra-Murillo}}, \bibinfo {author} {\bibfnamefont {P.~T.}\ \bibnamefont
  {Valentim}}, \bibinfo {author} {\bibfnamefont {J.~S.~V.}\ \bibnamefont
  {Morais}}, \bibinfo {author} {\bibfnamefont {F.}~\bibnamefont {Plentz}},
  \bibinfo {author} {\bibfnamefont {P.~S. S.~G.}\ \bibnamefont {{a}es}},
  \bibinfo {author} {\bibfnamefont {H.}~\bibnamefont {Vinck-Posada}}, \bibinfo
  {author} {\bibfnamefont {B.~A.}\ \bibnamefont {Rodriguez}}, \bibinfo {author}
  {\bibfnamefont {M.~S.}\ \bibnamefont {Skolnick}}, \bibinfo {author}
  {\bibfnamefont {A.}~\bibnamefont {Tahraoui}},\ and\ \bibinfo {author}
  {\bibfnamefont {M.}~\bibnamefont {Hopkinson}},\ }\bibfield  {title} {\bibinfo
  {title} {Quantum dot dipole orientation and excitation efficiency of
  micropillar modes},\ }\href {https://doi.org/10.1364/OE.16.019201} {\bibfield
   {journal} {\bibinfo  {journal} {Opt. Express}\ }\textbf {\bibinfo {volume}
  {16}},\ \bibinfo {pages} {19201} (\bibinfo {year} {2008})}\BibitemShut
  {NoStop}%
\bibitem [{\citenamefont {Usman}\ \emph {et~al.}(2009)\citenamefont {Usman},
  \citenamefont {Ryu}, \citenamefont {Woo}, \citenamefont {Ebert},\ and\
  \citenamefont {Klimeck}}]{4749336}%
  \BibitemOpen
  \bibfield  {author} {\bibinfo {author} {\bibfnamefont {M.}~\bibnamefont
  {Usman}}, \bibinfo {author} {\bibfnamefont {H.}~\bibnamefont {Ryu}}, \bibinfo
  {author} {\bibfnamefont {I.}~\bibnamefont {Woo}}, \bibinfo {author}
  {\bibfnamefont {D.~S.}\ \bibnamefont {Ebert}},\ and\ \bibinfo {author}
  {\bibfnamefont {G.}~\bibnamefont {Klimeck}},\ }\bibfield  {title} {\bibinfo
  {title} {Moving toward nano-tcad through multimillion-atom quantum-dot
  simulations matching experimental data},\ }\href
  {https://doi.org/10.1109/TNANO.2008.2011900} {\bibfield  {journal} {\bibinfo
  {journal} {IEEE Transactions on Nanotechnology}\ }\textbf {\bibinfo {volume}
  {8}},\ \bibinfo {pages} {330} (\bibinfo {year} {2009})}\BibitemShut {NoStop}%
\bibitem [{\citenamefont {Cortez}\ \emph {et~al.}(2001)\citenamefont {Cortez},
  \citenamefont {Krebs}, \citenamefont {Voisin},\ and\ \citenamefont
  {G\'erard}}]{PhysRevB.63.233306}%
  \BibitemOpen
  \bibfield  {author} {\bibinfo {author} {\bibfnamefont {S.}~\bibnamefont
  {Cortez}}, \bibinfo {author} {\bibfnamefont {O.}~\bibnamefont {Krebs}},
  \bibinfo {author} {\bibfnamefont {P.}~\bibnamefont {Voisin}},\ and\ \bibinfo
  {author} {\bibfnamefont {J.~M.}\ \bibnamefont {G\'erard}},\ }\bibfield
  {title} {\bibinfo {title} {Polarization of the interband optical dipole in
  {InAs/GaAs} self-organized quantum dots},\ }\href
  {https://doi.org/10.1103/PhysRevB.63.233306} {\bibfield  {journal} {\bibinfo
  {journal} {Phys. Rev. B}\ }\textbf {\bibinfo {volume} {63}},\ \bibinfo
  {pages} {233306} (\bibinfo {year} {2001})}\BibitemShut {NoStop}%
\bibitem [{\citenamefont {Qian}\ \emph {et~al.}(2019)\citenamefont {Qian},
  \citenamefont {Xie}, \citenamefont {Yang}, \citenamefont {Peng},
  \citenamefont {Wu}, \citenamefont {Song}, \citenamefont {Sun}, \citenamefont
  {Dang}, \citenamefont {Yu}, \citenamefont {Steer}, \citenamefont {Thayne},
  \citenamefont {Jin}, \citenamefont {Gu},\ and\ \citenamefont
  {Xu}}]{PhysRevLett.122.087401}%
  \BibitemOpen
  \bibfield  {author} {\bibinfo {author} {\bibfnamefont {C.}~\bibnamefont
  {Qian}}, \bibinfo {author} {\bibfnamefont {X.}~\bibnamefont {Xie}}, \bibinfo
  {author} {\bibfnamefont {J.}~\bibnamefont {Yang}}, \bibinfo {author}
  {\bibfnamefont {K.}~\bibnamefont {Peng}}, \bibinfo {author} {\bibfnamefont
  {S.}~\bibnamefont {Wu}}, \bibinfo {author} {\bibfnamefont {F.}~\bibnamefont
  {Song}}, \bibinfo {author} {\bibfnamefont {S.}~\bibnamefont {Sun}}, \bibinfo
  {author} {\bibfnamefont {J.}~\bibnamefont {Dang}}, \bibinfo {author}
  {\bibfnamefont {Y.}~\bibnamefont {Yu}}, \bibinfo {author} {\bibfnamefont
  {M.~J.}\ \bibnamefont {Steer}}, \bibinfo {author} {\bibfnamefont {I.~G.}\
  \bibnamefont {Thayne}}, \bibinfo {author} {\bibfnamefont {K.}~\bibnamefont
  {Jin}}, \bibinfo {author} {\bibfnamefont {C.}~\bibnamefont {Gu}},\ and\
  \bibinfo {author} {\bibfnamefont {X.}~\bibnamefont {Xu}},\ }\bibfield
  {title} {\bibinfo {title} {Enhanced strong interaction between nanocavities
  and $p$-shell excitons beyond the dipole approximation},\ }\href
  {https://doi.org/10.1103/PhysRevLett.122.087401} {\bibfield  {journal}
  {\bibinfo  {journal} {Phys. Rev. Lett.}\ }\textbf {\bibinfo {volume} {122}},\
  \bibinfo {pages} {087401} (\bibinfo {year} {2019})}\BibitemShut {NoStop}%
\bibitem [{\citenamefont {Li}\ \emph {et~al.}(2014)\citenamefont {Li},
  \citenamefont {Clements}, \citenamefont {Yu}, \citenamefont {Shi},
  \citenamefont {Gong},\ and\ \citenamefont
  {Xiao}}]{doi:10.1073/pnas.1408453111}%
  \BibitemOpen
  \bibfield  {author} {\bibinfo {author} {\bibfnamefont {B.-B.}\ \bibnamefont
  {Li}}, \bibinfo {author} {\bibfnamefont {W.~R.}\ \bibnamefont {Clements}},
  \bibinfo {author} {\bibfnamefont {X.-C.}\ \bibnamefont {Yu}}, \bibinfo
  {author} {\bibfnamefont {K.}~\bibnamefont {Shi}}, \bibinfo {author}
  {\bibfnamefont {Q.}~\bibnamefont {Gong}},\ and\ \bibinfo {author}
  {\bibfnamefont {Y.-F.}\ \bibnamefont {Xiao}},\ }\bibfield  {title} {\bibinfo
  {title} {Single nanoparticle detection using split-mode microcavity {Raman}
  lasers},\ }\href {https://doi.org/10.1073/pnas.1408453111} {\bibfield
  {journal} {\bibinfo  {journal} {Proc. Natl. Acad. Sci. U. S. A.}\ }\textbf
  {\bibinfo {volume} {111}},\ \bibinfo {pages} {14657} (\bibinfo {year}
  {2014})}\BibitemShut {NoStop}%
\bibitem [{\citenamefont {Yang}\ \emph {et~al.}(2020)\citenamefont {Yang},
  \citenamefont {Qian}, \citenamefont {Xie}, \citenamefont {Peng},
  \citenamefont {Wu}, \citenamefont {Song}, \citenamefont {Sun}, \citenamefont
  {Dang}, \citenamefont {Yu}, \citenamefont {Shi}, \citenamefont {He},
  \citenamefont {Steer}, \citenamefont {Thayne}, \citenamefont {Li},
  \citenamefont {Bo}, \citenamefont {Xiao}, \citenamefont {Zuo}, \citenamefont
  {Jin}, \citenamefont {Gu},\ and\ \citenamefont {Xu}}]{Yang2020}%
  \BibitemOpen
  \bibfield  {author} {\bibinfo {author} {\bibfnamefont {J.}~\bibnamefont
  {Yang}}, \bibinfo {author} {\bibfnamefont {C.}~\bibnamefont {Qian}}, \bibinfo
  {author} {\bibfnamefont {X.}~\bibnamefont {Xie}}, \bibinfo {author}
  {\bibfnamefont {K.}~\bibnamefont {Peng}}, \bibinfo {author} {\bibfnamefont
  {S.}~\bibnamefont {Wu}}, \bibinfo {author} {\bibfnamefont {F.}~\bibnamefont
  {Song}}, \bibinfo {author} {\bibfnamefont {S.}~\bibnamefont {Sun}}, \bibinfo
  {author} {\bibfnamefont {J.}~\bibnamefont {Dang}}, \bibinfo {author}
  {\bibfnamefont {Y.}~\bibnamefont {Yu}}, \bibinfo {author} {\bibfnamefont
  {S.}~\bibnamefont {Shi}}, \bibinfo {author} {\bibfnamefont {J.}~\bibnamefont
  {He}}, \bibinfo {author} {\bibfnamefont {M.~J.}\ \bibnamefont {Steer}},
  \bibinfo {author} {\bibfnamefont {I.~G.}\ \bibnamefont {Thayne}}, \bibinfo
  {author} {\bibfnamefont {B.-B.}\ \bibnamefont {Li}}, \bibinfo {author}
  {\bibfnamefont {F.}~\bibnamefont {Bo}}, \bibinfo {author} {\bibfnamefont
  {Y.-F.}\ \bibnamefont {Xiao}}, \bibinfo {author} {\bibfnamefont
  {Z.}~\bibnamefont {Zuo}}, \bibinfo {author} {\bibfnamefont {K.}~\bibnamefont
  {Jin}}, \bibinfo {author} {\bibfnamefont {C.}~\bibnamefont {Gu}},\ and\
  \bibinfo {author} {\bibfnamefont {X.}~\bibnamefont {Xu}},\ }\bibfield
  {title} {\bibinfo {title} {Diabolical points in coupled active cavities with
  quantum emitters},\ }\href {https://doi.org/10.1038/s41377-020-0244-9}
  {\bibfield  {journal} {\bibinfo  {journal} {Light: Sci. Appl.}\ }\textbf
  {\bibinfo {volume} {9}},\ \bibinfo {pages} {6} (\bibinfo {year}
  {2020})}\BibitemShut {NoStop}%
\bibitem [{\citenamefont {Kippenberg}\ \emph {et~al.}(2002)\citenamefont
  {Kippenberg}, \citenamefont {Spillane},\ and\ \citenamefont
  {Vahala}}]{Kippenberg:02}%
  \BibitemOpen
  \bibfield  {author} {\bibinfo {author} {\bibfnamefont {T.~J.}\ \bibnamefont
  {Kippenberg}}, \bibinfo {author} {\bibfnamefont {S.~M.}\ \bibnamefont
  {Spillane}},\ and\ \bibinfo {author} {\bibfnamefont {K.~J.}\ \bibnamefont
  {Vahala}},\ }\bibfield  {title} {\bibinfo {title} {Modal coupling in
  traveling-wave resonators},\ }\href {https://doi.org/10.1364/OL.27.001669}
  {\bibfield  {journal} {\bibinfo  {journal} {Opt. Lett.}\ }\textbf {\bibinfo
  {volume} {27}},\ \bibinfo {pages} {1669} (\bibinfo {year}
  {2002})}\BibitemShut {NoStop}%
\bibitem [{\citenamefont {Li}\ \emph {et~al.}(2012)\citenamefont {Li},
  \citenamefont {Eftekhar}, \citenamefont {Xia},\ and\ \citenamefont
  {Adibi}}]{Li:12}%
  \BibitemOpen
  \bibfield  {author} {\bibinfo {author} {\bibfnamefont {Q.}~\bibnamefont
  {Li}}, \bibinfo {author} {\bibfnamefont {A.~A.}\ \bibnamefont {Eftekhar}},
  \bibinfo {author} {\bibfnamefont {Z.}~\bibnamefont {Xia}},\ and\ \bibinfo
  {author} {\bibfnamefont {A.}~\bibnamefont {Adibi}},\ }\bibfield  {title}
  {\bibinfo {title} {Azimuthal-order variations of surface-roughness-induced
  mode splitting and scattering loss in {high-Q} microdisk resonators},\ }\href
  {https://doi.org/10.1364/OL.37.001586} {\bibfield  {journal} {\bibinfo
  {journal} {Opt. Lett.}\ }\textbf {\bibinfo {volume} {37}},\ \bibinfo {pages}
  {1586} (\bibinfo {year} {2012})}\BibitemShut {NoStop}%
\bibitem [{\citenamefont {Moiseyev}\ and\ \citenamefont
  {Friedland}(1980)}]{PhysRevA.22.618}%
  \BibitemOpen
  \bibfield  {author} {\bibinfo {author} {\bibfnamefont {N.}~\bibnamefont
  {Moiseyev}}\ and\ \bibinfo {author} {\bibfnamefont {S.}~\bibnamefont
  {Friedland}},\ }\bibfield  {title} {\bibinfo {title} {Association of
  resonance states with the incomplete spectrum of finite complex-scaled
  hamiltonian matrices},\ }\href {https://doi.org/10.1103/PhysRevA.22.618}
  {\bibfield  {journal} {\bibinfo  {journal} {Phys. Rev. A}\ }\textbf {\bibinfo
  {volume} {22}},\ \bibinfo {pages} {618} (\bibinfo {year} {1980})}\BibitemShut
  {NoStop}%
\bibitem [{\citenamefont {Zhu}\ \emph {et~al.}(2020)\citenamefont {Zhu},
  \citenamefont {Liu}, \citenamefont {Bo}, \citenamefont {Tao}, \citenamefont
  {Zhang},\ and\ \citenamefont {Xu}}]{PhysRevA.101.053842}%
  \BibitemOpen
  \bibfield  {author} {\bibinfo {author} {\bibfnamefont {J.}~\bibnamefont
  {Zhu}}, \bibinfo {author} {\bibfnamefont {H.}~\bibnamefont {Liu}}, \bibinfo
  {author} {\bibfnamefont {F.}~\bibnamefont {Bo}}, \bibinfo {author}
  {\bibfnamefont {C.}~\bibnamefont {Tao}}, \bibinfo {author} {\bibfnamefont
  {G.}~\bibnamefont {Zhang}},\ and\ \bibinfo {author} {\bibfnamefont
  {J.}~\bibnamefont {Xu}},\ }\bibfield  {title} {\bibinfo {title} {Intuitive
  model of exceptional points in an optical whispering-gallery microcavity
  perturbed by nanoparticles},\ }\href
  {https://doi.org/10.1103/PhysRevA.101.053842} {\bibfield  {journal} {\bibinfo
   {journal} {Phys. Rev. A}\ }\textbf {\bibinfo {volume} {101}},\ \bibinfo
  {pages} {053842} (\bibinfo {year} {2020})}\BibitemShut {NoStop}%
\bibitem [{\citenamefont {Yuan}\ \emph {et~al.}(2018)\citenamefont {Yuan},
  \citenamefont {Lin}, \citenamefont {Xiao},\ and\ \citenamefont
  {Fan}}]{Yuan:18}%
  \BibitemOpen
  \bibfield  {author} {\bibinfo {author} {\bibfnamefont {L.}~\bibnamefont
  {Yuan}}, \bibinfo {author} {\bibfnamefont {Q.}~\bibnamefont {Lin}}, \bibinfo
  {author} {\bibfnamefont {M.}~\bibnamefont {Xiao}},\ and\ \bibinfo {author}
  {\bibfnamefont {S.}~\bibnamefont {Fan}},\ }\bibfield  {title} {\bibinfo
  {title} {Synthetic dimension in photonics},\ }\href
  {https://doi.org/10.1364/OPTICA.5.001396} {\bibfield  {journal} {\bibinfo
  {journal} {Optica}\ }\textbf {\bibinfo {volume} {5}},\ \bibinfo {pages}
  {1396} (\bibinfo {year} {2018})}\BibitemShut {NoStop}%
\bibitem [{\citenamefont {Zhou}\ \emph {et~al.}(2017)\citenamefont {Zhou},
  \citenamefont {Luo}, \citenamefont {Wang}, \citenamefont {Guo}, \citenamefont
  {Zhou}, \citenamefont {Pu},\ and\ \citenamefont
  {Zhou}}]{PhysRevLett.118.083603}%
  \BibitemOpen
  \bibfield  {author} {\bibinfo {author} {\bibfnamefont {X.-F.}\ \bibnamefont
  {Zhou}}, \bibinfo {author} {\bibfnamefont {X.-W.}\ \bibnamefont {Luo}},
  \bibinfo {author} {\bibfnamefont {S.}~\bibnamefont {Wang}}, \bibinfo {author}
  {\bibfnamefont {G.-C.}\ \bibnamefont {Guo}}, \bibinfo {author} {\bibfnamefont
  {X.}~\bibnamefont {Zhou}}, \bibinfo {author} {\bibfnamefont {H.}~\bibnamefont
  {Pu}},\ and\ \bibinfo {author} {\bibfnamefont {Z.-W.}\ \bibnamefont {Zhou}},\
  }\bibfield  {title} {\bibinfo {title} {Dynamically manipulating topological
  physics and edge modes in a single degenerate optical cavity},\ }\href
  {https://doi.org/10.1103/PhysRevLett.118.083603} {\bibfield  {journal}
  {\bibinfo  {journal} {Phys. Rev. Lett.}\ }\textbf {\bibinfo {volume} {118}},\
  \bibinfo {pages} {083603} (\bibinfo {year} {2017})}\BibitemShut {NoStop}%
\bibitem [{\citenamefont {Yuan}\ \emph {et~al.}(2019)\citenamefont {Yuan},
  \citenamefont {Lin}, \citenamefont {Zhang}, \citenamefont {Xiao},
  \citenamefont {Chen},\ and\ \citenamefont {Fan}}]{PhysRevLett.122.083903}%
  \BibitemOpen
  \bibfield  {author} {\bibinfo {author} {\bibfnamefont {L.}~\bibnamefont
  {Yuan}}, \bibinfo {author} {\bibfnamefont {Q.}~\bibnamefont {Lin}}, \bibinfo
  {author} {\bibfnamefont {A.}~\bibnamefont {Zhang}}, \bibinfo {author}
  {\bibfnamefont {M.}~\bibnamefont {Xiao}}, \bibinfo {author} {\bibfnamefont
  {X.}~\bibnamefont {Chen}},\ and\ \bibinfo {author} {\bibfnamefont
  {S.}~\bibnamefont {Fan}},\ }\bibfield  {title} {\bibinfo {title} {Photonic
  gauge potential in one cavity with synthetic frequency and orbital angular
  momentum dimensions},\ }\href
  {https://doi.org/10.1103/PhysRevLett.122.083903} {\bibfield  {journal}
  {\bibinfo  {journal} {Phys. Rev. Lett.}\ }\textbf {\bibinfo {volume} {122}},\
  \bibinfo {pages} {083903} (\bibinfo {year} {2019})}\BibitemShut {NoStop}%
\bibitem [{\citenamefont {Dutt}\ \emph {et~al.}(2020)\citenamefont {Dutt},
  \citenamefont {Lin}, \citenamefont {Yuan}, \citenamefont {Minkov},
  \citenamefont {Xiao},\ and\ \citenamefont
  {Fan}}]{doi:10.1126/science.aaz3071}%
  \BibitemOpen
  \bibfield  {author} {\bibinfo {author} {\bibfnamefont {A.}~\bibnamefont
  {Dutt}}, \bibinfo {author} {\bibfnamefont {Q.}~\bibnamefont {Lin}}, \bibinfo
  {author} {\bibfnamefont {L.}~\bibnamefont {Yuan}}, \bibinfo {author}
  {\bibfnamefont {M.}~\bibnamefont {Minkov}}, \bibinfo {author} {\bibfnamefont
  {M.}~\bibnamefont {Xiao}},\ and\ \bibinfo {author} {\bibfnamefont
  {S.}~\bibnamefont {Fan}},\ }\bibfield  {title} {\bibinfo {title} {A single
  photonic cavity with two independent physical synthetic dimensions},\ }\href
  {https://doi.org/10.1126/science.aaz3071} {\bibfield  {journal} {\bibinfo
  {journal} {Science}\ }\textbf {\bibinfo {volume} {367}},\ \bibinfo {pages}
  {59} (\bibinfo {year} {2020})}\BibitemShut {NoStop}%
\bibitem [{\citenamefont {Liu}\ \emph {et~al.}(2019)\citenamefont {Liu},
  \citenamefont {Wang}, \citenamefont {Zhang}, \citenamefont {Wang},
  \citenamefont {Zhao}, \citenamefont {Guan}, \citenamefont {Liu},
  \citenamefont {Shi},\ and\ \citenamefont {Zi}}]{PhysRevLett.123.116104}%
  \BibitemOpen
  \bibfield  {author} {\bibinfo {author} {\bibfnamefont {W.}~\bibnamefont
  {Liu}}, \bibinfo {author} {\bibfnamefont {B.}~\bibnamefont {Wang}}, \bibinfo
  {author} {\bibfnamefont {Y.}~\bibnamefont {Zhang}}, \bibinfo {author}
  {\bibfnamefont {J.}~\bibnamefont {Wang}}, \bibinfo {author} {\bibfnamefont
  {M.}~\bibnamefont {Zhao}}, \bibinfo {author} {\bibfnamefont {F.}~\bibnamefont
  {Guan}}, \bibinfo {author} {\bibfnamefont {X.}~\bibnamefont {Liu}}, \bibinfo
  {author} {\bibfnamefont {L.}~\bibnamefont {Shi}},\ and\ \bibinfo {author}
  {\bibfnamefont {J.}~\bibnamefont {Zi}},\ }\bibfield  {title} {\bibinfo
  {title} {Circularly polarized states spawning from bound states in the
  continuum},\ }\href {https://doi.org/10.1103/PhysRevLett.123.116104}
  {\bibfield  {journal} {\bibinfo  {journal} {Phys. Rev. Lett.}\ }\textbf
  {\bibinfo {volume} {123}},\ \bibinfo {pages} {116104} (\bibinfo {year}
  {2019})}\BibitemShut {NoStop}%
\end{thebibliography}
\end{document}